\documentclass[
twocolumn,
nofootinbib,
 amsmath,amssymb,
 aps,
prb,
superscriptaddress,
10pt
]{revtex4-2}
\usepackage{chemformula}
\usepackage{graphicx}
\usepackage{dcolumn}
\usepackage{comment}
\usepackage[normalem]{ulem}
\usepackage{bm}
\usepackage{hyperref}
\usepackage{tikz}
\usepackage{xcolor}
\usepackage{pdfpages}
\usepackage{pgffor}

\makeatletter
\AtBeginDocument{\let\LS@rot\@undefined}
\makeatother

\begin{document}


\title{Score-based diffusion models for accurate crystal-structure inpainting and reconstruction of hydrogen positions}

\author{Timo Reents}
\affiliation{
PSI Center for Scientific Computing, Theory and Data, Paul Scherrer Institute, 5232 Villigen PSI, Switzerland
}%

\author{Arianna Cantarella}
\affiliation{
Dipartimento di Scienze Matematiche, Fisiche e Informatiche, Università di Parma, I-43124 Parma, Italy
}%
\affiliation{INFN - Sezione di Milano-Bicocca, gruppo collegato di Parma, 43124 Parma, Italy}

\author{Marnik Bercx}%
\affiliation{
PSI Center for Scientific Computing, Theory and Data, Paul Scherrer Institute, 5232 Villigen PSI, Switzerland
}%

\author{Pietro Bonf\`a}
\affiliation{
Dipartimento di Scienze Fisiche, Informatiche e Matematiche, Università degli Studi di Modena e Reggio Emilia, Modena, Italy
}%
\affiliation{Institute Nanoscience - CNR-NANO, Center S3, Modena, Italy}

\author{Giovanni Pizzi}
\email{Corresponding author: giovanni.pizzi@psi.ch}
\affiliation{
PSI Center for Scientific Computing, Theory and Data, Paul Scherrer Institute, 5232 Villigen PSI, Switzerland
}%

\begin{abstract}
Generative artificial-intelligence (AI) models, such as score-based diffusion models, have recently advanced the field of computational materials science by enabling the generation of new materials with desired properties. In addition, these models could also be leveraged to reconstruct crystal structures for which partial information is available. 
One relevant example is the reliable determination of atomic positions occupied by hydrogen atoms in hydrogen-containing crystalline materials.
While crucial to the analysis and prediction of many materials properties, the identification of hydrogen positions via X-ray scattering experiments has been historically challenging, and often required more expensive neutron scattering measurements. As a consequence, and despite experimental advances which enable nowadays to accurately determine hydrogen positions based on X-ray scattering experiments, inorganic crystallographic databases still report many lattice structures where hydrogen atoms have been either omitted or inserted with heuristics or by chemical intuition. Here, we combine diffusion models from the field of materials science with techniques originally developed in computer vision for image inpainting. We present how this knowledge transfer across domains enables a much faster and more accurate completion of host structures, compared to unconditioned diffusion models or previous approaches solely based on density-functional theory (DFT). Overall, when applied to a test dataset of hydrogen-containing materials from the MC3D database, our approach exceeds a success rate of 97\% in terms of finding a structural match or predicting a more stable configuration (according to DFT) than the initial reference from the experimental source database (and with a success rate exceeding 99\% when excluding structures flagged as theoretical in MC3D), both when starting from structures that were already relaxed with DFT, or when starting directly from the experimentally determined host structures.

\end{abstract}

\maketitle

\begin{figure*}

\includegraphics[width=\linewidth]{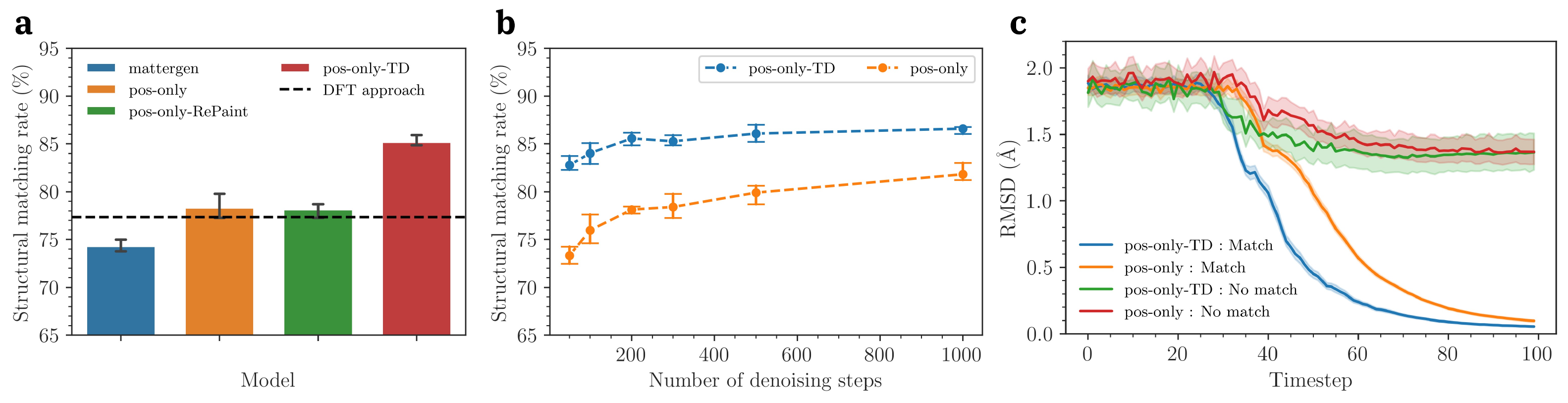}
\caption{
\textbf{Comparison of the single-trial performance of different inpainting approaches.}
\textbf{a} Structural matching rate for multiple models and inpainting approaches: \texttt{MatterGen}, \texttt{pos‑only}, \texttt{pos-only-RePaint} and the \texttt{pos‑only-TD}, see main text for description of the models. Furthermore, we also show the performance of our DFT-based reconstruction approach discussed in the SI section~S2 (dashed line).
\textbf{b} Structural matching rate as a function of the denoising steps for the \texttt{pos‑only-TD} and \texttt{pos‑only} model.
\textbf{c} Root mean square deviation (RMSD) of the predictions along the diffusion trajectory with respect to the reference structure, normalized by the number of hydrogen sites. The results are shown for the \texttt{pos‑only-TD} and \texttt{pos‑only} model and are distinguished into those cases that result in a final structural match and those that do not.
The error bars in \textbf{a} and \textbf{b} result from 4 runs for each datapoint and show the min-max range of the single-trial structural matching rates.
}
\label{fig:DL:compare-inpainting-methods}
\end{figure*}

\section{Introduction}
The field of crystal structure prediction (CSP) recently experienced significant progress with the emergence of deep-learning based generative models~\cite{ryan_crystal_2018,cheng_crystal_2022,xie_crystal_2022,park_has_2024, podryabinkin_accelerating_2019, kim_generative_2020, zhao_physics_2023}. In particular, score-based diffusion models have become highly popular in many domains~\cite{song_generative_2020,song_score-based_2021,zeni_generative_2025,levy_symmcd_2025,jiao_crystal_2024,luo_deep_2024,li_generative_2025,pakornchote_diffusion_2024}.
These are used to learn and sample a complex distribution by adding and removing (typically) Gaussian noise according to a stochastic differential equation~\cite{song_score-based_2021}. While most of these models, and also traditional CSP efforts prior to the deep learning era~\cite{oganov_how_2011, woodley_crystal_2008}, focus on the generation of stable crystal structures given the target composition~\cite{breuck_generative_2025}, another very important challenge is the reconstruction and completion of crystal structures for which only partial information is available~\cite{dai2024inpainting}. Such applications range from intercalating ions into potential cathode materials to materials for which the experimental determination of the crystal structure is challenging.

One example belonging to the latter category are crystalline materials that contain hydrogen. 
Hydrogen is one of the most abundant elements on our planet.
In addition to its fundamental contribution to life, it is a promising carbon-neutral energy carrier~\cite{mazloomi_hydrogen_2012,ge_review_2024}, it plays a key role in semiconductor industry \cite{10.1146/annurev.matsci.36.010705.155428}, and it is intensively studied in many other fields ranging from photocatalysis~\cite{guo_boosting_2021, christoforidis_photocatalytic_2017} to fusion reactors \cite{ROTH20091}.
Unfortunately, hydrogen has the lowest X-ray and electron scattering power of all atoms and, at the same time, it has one of the highest incoherent scattering cross sections of all isotopes, which represents an unwanted background contribution for structural refinements with neutron diffraction.
Nevertheless, neutron diffraction is the experimental method that sets and has set the reference for accurate identification of H positions~\cite{Machida2014}. However, this technique requires large facilities and substantial amounts of sample, thus making these experiments orders of magnitude more expensive than standard laboratory X-ray diffraction. 
Similarly, synchrotron-radiation X-ray powder diffraction has been applied in the context of hydrogen storage materials~\cite{Cheng2017}, but again this requires the availability of large facilities.
Other techniques like 3D electron diffraction, based on the dynamical diffraction theory (instead of the kinematical approximation of single scattering)~\cite{ED-dyn-refine-Palatinus2015,ED-Palatinus2015-theory}, are also becoming increasingly popular as alternatives to X-ray diffraction methods, especially when the crystal sizes are limiting the application of single-crystal X-ray diffraction. Their application has successfully been proven for the determination of crystal structures in general~\cite{ED-Gemmi2019}, but also for hydrogen containing compounds~\cite{Palatinus2017, ED-Kumar2024,ED-Klar2023}.
Finally, thanks to the developments in the field of quantum crystallography~\cite{QC-Grabowsky2017,QC-Genoni2020}, the applicability of X-ray diffraction methods to locate hydrogen atoms in materials has greatly improved, based on approaches such as Hirshfeld atom refinement~\cite{Hirshfeld-Jayatilaka2008, Hirshfeld-Capelli2014} (i.e., using non-spherical electron densities instead of the independent atom model), almost reaching the accuracy of neutron diffraction~\cite{sciadv.1600192, Hirshfeld-Woiska2021, Kleemiss2021} and, at the same time, these methods have become much more widely applicable~\cite{Kleemiss2021}.

Despite these experimental advances that enable the accurate location of hydrogen positions in materials, to date, the available (historical) data in experimental databases often report lattice structures nominally containing H atoms, with hydrogen positions that have either been introduced with heuristics (it is sometimes stated that H atoms are placed into ``chemically sensible positions'') or do not provide H positions at all~\cite{Cooper2010,Afonine2010}. Moreover, the quality of the data can be heterogeneous across entries, especially when the uncertainty is large or the data was obtained for a specific purpose, with a different focus than accurately determining the crystal structure~\cite{Spek:su5533}. This is particularly severe and relevant in the context of materials databases~\mbox{\cite{jain_commentary_2013,kirklin_open_2015,schmidt_dataset_2022,barroso-luque_open_2024,horton_accelerated_2025,kaplan_foundational_2025,huber_mc3d_2025}} that rely on the availability of complete and curated crystal structures, such as those that employ atomistic models to predict their properties.

In this work, we directly leverage and retrain a new version of Microsoft's \texttt{MatterGen} model~\cite{zeni_generative_2025}, a diffusion model originally trained to generate stable, unique and novel crystal structures with desired properties~\cite{chen_accelerating_2025, park_guiding_2025}, to efficiently reconstruct crystal structures with partial information available. We first show how the domain of materials science can benefit from methods originally developed in the context of image inpainting in computer vision tasks~\cite{lugmayr_repaint_2022,mayet_td-paint_2024}. Specifically, we choose the task of finding missing hydrogen positions in materials, due to the challenges related to their experimental measurement, as outlined before. By combining the diffusion model with recent advances in the development of machine learning interatomic potentials MLIPs~\cite{riebesell_framework_2025,yang_mattersim_2024,batzner_e3-equivariant_2022,batatia_foundation_2024,rhodes_orb-v3_2025,yu_systematic_2024,mazitov_pet-mad_2025} to drive structural optimization, we achieve success rates greater than 97\% in terms of structurally matching our original reference or predicting configurations that are actually more stable than the original reference. Thanks to this high success rate, the method that we present here can also be used to predict new candidate structures for hydrogen related research, which can then be screened for interesting and desired properties. Importantly, the models we present are trained on top of \texttt{MatterGen} in a hydrogen-agnostic way, meaning that the method and models can be easily transferred to other related tasks, such as intercalation, without the necessity of retraining.

\section{Results}

\subsection{Comparison of inpainting approaches}
\label{ssec:results-inpainting-comparison}
We start by demonstrating how our methodological advances can efficiently and accurately perform crystal structure inpainting.
In this first section, we will only focus on actual structural matches, meaning whether our approach is capable of correctly restoring the hydrogen positions with respect to the reference, as determined by \texttt{pymatgen}'s \texttt{StructureMatcher}. Moreover, in this section, we will only generate one trial per structure.
While this already achieves a notable performance, we will show later on in section~\ref{ssec:results-LES-performance} how the performance can be increased even more, yielding the aforementioned 97\% success rate.
This is achieved by generating several trials per structure and by considering ``energetic matches'', i.e., predictions that find a lower energy configuration given the host structure (i.e., the original reference structure with the H atoms removed) compared to original reference.

Fig.~\ref{fig:DL:compare-inpainting-methods}a compares different inpainting approaches and models, that are described in more detail in the Methods section~\ref{ssec:models-and-training}: the \texttt{MatterGen} baseline model (from reference~\cite{zeni_generative_2025}, originally trained to denoise the positions, lattice and atomic types, even though the lattice and atomic types are fixed for the denoising in these inpainting experiments),
\texttt{pos-only} (our newly (re)trained model to only denoise the positions), \texttt{pos-only-RePaint} (the previous newly trained model, but evaluated in combination with the \texttt{RePaint} algorithm~\cite{lugmayr_repaint_2022}) 
and the \texttt{pos-only-TD} model (our newly trained model with different noise levels per structure, transferring the concept of \texttt{TD-Paint}~\cite{mayet_td-paint_2024} from computer vision to materials science). 

As general note, as this will be relevant in the context of this section, we remind the reader that score-based diffusion models are generative models that are trained to iteratively remove noise from samples (typically starting from Gaussian noise), to transform the initial noisy samples into ``clean'' samples that follow the target distribution.
We will call the number of denoising steps $N_{steps}$ in the following. Moreover, we briefly introduce the aforementioned inpainting approaches \texttt{RePaint} and \texttt{TD-Paint}, originally developed for image inpainting (more details can be found in the Methods section).
\texttt{RePaint} is an inpainting approach applicable to any unconditional diffusion-based model (without retraining) and improves the inpainting performance by repeatedly going back and forth in the denoising process. By resampling the known pixels during the denoising process, this approach harmonizes the inpainted region with the known region more effectively.
\texttt{TD-Paint} is a strategy to (re-)train a diffusion-based model that handles variable noise levels at the pixel level, instead of a uniform noise level per image, thus allowing to directly condition its denoising process on the available image information, keeping known regions fixed (no noise) while iteratively predicting the missing parts. As a result, missing regions can be reconstructed more efficiently by tailoring the noise update to the known context.

We benchmark the inpainting approaches on a \texttt{DFT} dataset which consists of a set of inorganic crystal structures (with known hydrogen positions) that were curated by means of DFT and taken from the MC3D database~\cite{huber_mc3d_2025} (version \texttt{PBE-v1}). More details on the datasets can be found in the Methods section~\ref{ssec:models-and-training}. For the rest of this work, we remove the hydrogen sites to generate host structures, then predict back the H position using our method (here: diffusion model + constrained relaxation of hydrogen positions, see Fig.~S1 for the impact of the constrained relaxation) and compare them against the initial known references. 

It is clearly visible that the retraining of a model that only denoises the positions already improves the structural matching rate.
It is interesting to note that applying the \texttt{RePaint} algorithm on top of this retrained model only reduces the variance, but does not really impact the overall performance. However, the number of steps increases from 300 to around 1000, making the prediction computationally more expensive. Moreover, to really notice a significant advantage of the \texttt{RePaint} approach in this task, one needs to use even more steps, e.g. by also increasing the number of corrector steps~\cite{Zhong2025}, to around 4000. This is further discussed in the corresponding Methods section~\ref{ssec:methods-inpainting} and the SI Fig.~S2~and~S3.
Finally, our new \texttt{pos-only-TD} model yields the highest structural matching rate. 
This confirms the expectation that the model better ``understands'' the conditioning on the known positions, as it always conditions on the clean known positions instead of a noisy version of them (analogously to the original \texttt{TD-Paint} for known portions of an image). Finally, we also show the performance of a purely DFT-based approach that we implemented and use as a benchmark and reference (using the electrostatic potential to estimate the hydrogen positions), see Methods~\ref{ssec:DFT-reconstruction}, evaluated on a slightly smaller set of structures (only the common structures present in the \texttt{DFT} dataset and in the dataset used for the evaluation of the DFT-based reconstruction approach, more details in the SI section~S2), yielding a matching rate of $\approx 77\%$. As described in the Methods section, the DFT based approach might require the so-called \textit{pinball} method for certain structures~\cite{kahle_modeling_2018}. The matching rate increases to almost $87\%$ when neglecting those cases. However, we note that also the \texttt{pos-only-TD} model reaches a higher matching rate of $88\%$ on that subset. While these DFT results show the overall good performance of such a physics-inspired DFT-based approach, they also emphasize its limitations.
First, the DFT-based method has a higher computational cost, both due to the cost of running DFT simulations, and to the combinatorial challenge involved in the \textit{pinball} method~\cite{kahle_modeling_2018}. 
Second, as we discuss in detail in the SI section~S2, the performance is limited by the nature of the underlying greedy optimization algorithm (see SI section~S2\,E) and it is very difficult to further increase the success rate of the method. In the following, instead, we demonstrate how our data-driven score-based diffusion model approach can be seen as an efficient global optimization method and can achieve much higher success rates. 


In order to analyze whether the prediction process can be made more efficient, 
Fig.~\ref{fig:DL:compare-inpainting-methods}b presents the impact of the number of denoising steps on the structural matching rate. We note that the models were trained on 1000 denoising steps (i.e., noise levels). Due to the superior performance and lower computational cost of the \texttt{pos-only-TD} model over the other inpainting \texttt{pos-only-RePaint} model, we only focus on the \texttt{pos-only-TD} and \texttt{pos-only} models here. Again, it is clearly visible that the \texttt{pos-only-TD} model performs significantly better. In terms of computational efficiency, this analysis shows that the number of denoising steps could be reduced from 1000 to 300 steps during inference, almost without any reduction in performance. Notably, even with only 50 steps, the \texttt{pos-only-TD} model performs much better than the \texttt{pos-only} model (even with more steps).

Furthermore, the spread between the unrelaxed and relaxed predictions is much smaller in case of the \texttt{pos-only-TD} model compared to the one that is observed for the \texttt{pos-only} model (not shown but discussed in SI Fig.~S1), indicating that the \texttt{pos-only-TD} model generates configurations that are already closer to the local minimum and in better agreement with the known part of the host structure. 

\begin{figure*}
\includegraphics[width=\linewidth]{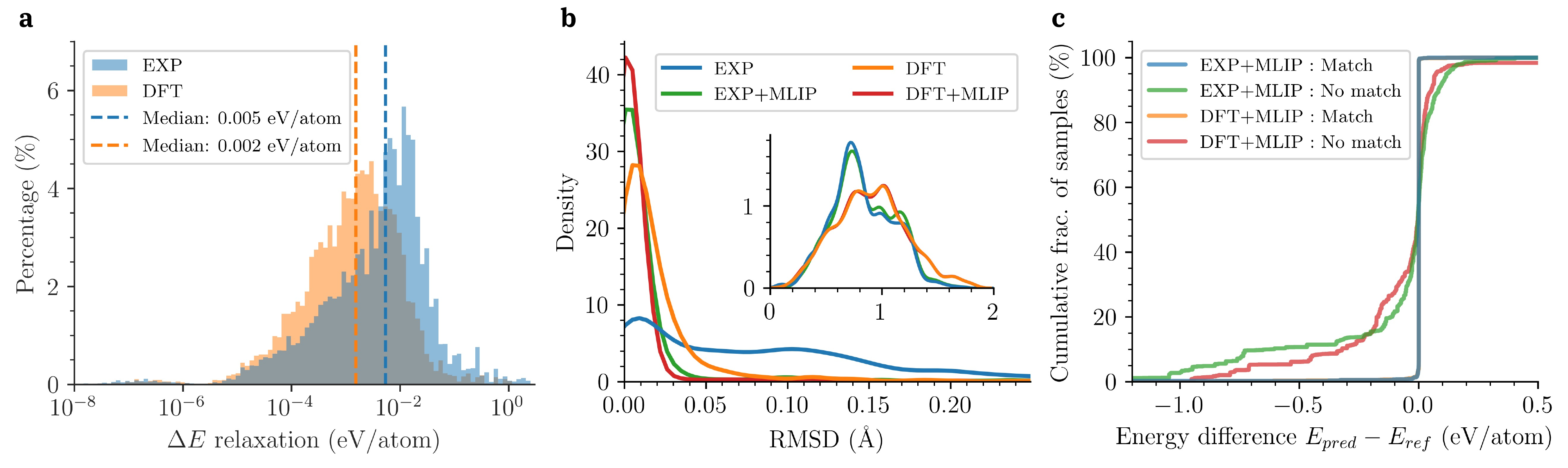}
\caption{\textbf{Analysis of the energetic and structural agreement of the predictions with respect to the references.}
\textbf{a} Only for the matching predictions: Difference between the energy of the initial structure generated by \texttt{MatterGen} and its energy after structural relaxation using \texttt{NequIP}. The color distinguishes between \texttt{DFT} and \texttt{EXP} datasets.
\textbf{b} Kernel density estimate (KDE) plot of the root mean square deviation (RMSD) between the atomic positions of the prediction and the references. The main plot shows the distribution for the samples that result in a structural match, the inset for the non-matching ones.
\textbf{c} Cumulative percentage of the energy difference between the prediction and reference (the reference has been relaxed with the MLIP as well, in order to have comparable energies and methods). We distinguish each of the datasets into two sub groups, those resulting in a structural match and those that do not. Negative energy differences indicate predicted structures that are more stable than the initial reference (see discussion in the main text).
}
\label{fig:DL:structural-energetic-agreement-MLIP}
\end{figure*}

To further analyze the denoising process and how it differs between the two models, Fig.~\ref{fig:DL:compare-inpainting-methods}c presents the root mean square deviation (RMSD) of the hydrogen positions along the whole trajectory with respect to the final reference structure (normalized by the number of hydrogen sites). Note that we use the mean at each step, i.e., the term without adding the Gaussian noise in Eq.~\eqref{eq:reverse-discretized}. While the \texttt{pos-only-TD} and \texttt{pos-only} model behave similarly in those cases for which the predictions do not match the reference, two main differences can be observed when focusing on the cases that result in a final structural match. 
First, the RMSD starts to decrease earlier and exhibits a more negative slope, indicating that the correct manifold is reached faster, in agreement with the underlying idea of the \texttt{TD-Paint} approach~\cite{mayet_td-paint_2024}. Second, the RMSD of the final prediction is lower in case of the \texttt{pos-only-TD} model, once again highlighting the advantage of the \texttt{pos-only-TD} approach.

The previous analyses clearly demonstrate the efficiency and superior performance of the \texttt{TD-Paint} approach to reconstruct sites in host structures. 
All of the following results are based on our \texttt{pos-only-TD} model, with the final workflow being structured as follows:
\vspace{\baselineskip}
\begin{enumerate}
    \item Start from a host structure and randomly initialize $N_H$ hydrogen sites in the unit cell. At this stage, it is assumed that the number of sites to be added is known. We refer to the discussion section~\ref{sec:discussion} for further remarks and SI Table~S1 in section~S1\,B for some proof of concept on how to estimate the number of missing sites.
    \item Apply the score-based diffusion model \texttt{pos-only-TD} for $N_{steps}$ iterations.
    \item Perform a constrained relaxation of the hydrogen sites using a MLIP (\texttt{NequIP}~\cite{kavanagh_2025_16980200} in this case; we discuss the choice of MLIP in the SI section~S1\,C, Fig.~S4-S6), keeping the cell and all remaining sites fixed.
    \item Perform a full relaxation of all positions, again using a MLIP (\texttt{NequIP} in this case), including the non-hydrogen sites.
\end{enumerate}

In addition to the \texttt{DFT} dataset, we now also evaluate our method on the \texttt{EXP} dataset, which consists of the underlying experimental structures (originally obtained from the experimental source databases COD~\cite{cod}, ICSD~\cite{icsd} and MPDS~\cite{mpds}) before the curation by DFT to construct the MC3D database, see section~\ref{ssec:models-and-training} for more details. This dataset is more challenging as \texttt{MatterGen} is trained on DFT data, but it is a representative benchmarking for actual application, where preliminary DFT relaxations (with known H positions) are clearly not available. We note here that certain structures are labeled as \textit{theoretical} in the source databases. 

Moreover, up to now, only one sample (i.e., one trial per structure) has been generated. 
Since a diffusion model follows a probabilistic approach, several samples per structure can be generated that, in turn, might produce several different solutions. 
We will discuss later how, if several distinct solutions are obtained, a comparison of their total energies (either with a MLIP or with DFT) can then help to identify the most stable one. 
To demonstrate how multiple samples improve the success rate of our method, we generate \mbox{30 samples} for each structure both in the \texttt{DFT} and \texttt{EXP} datasets, which we use for our subsequent analyses to compare the structural and energetic agreement of our predictions with respect to the reference data.

\subsection{Energetic and structural agreement of the predictions: identification of more stable structures}

We start with the discussion of the energy differences between the structures generated by the diffusion model and after relaxing them (i.e., comparing the outputs from step 2 and 4 in the workflow outlined before), shown in Fig.~\ref{fig:DL:structural-energetic-agreement-MLIP}a, both for the \texttt{DFT} and \texttt{EXP} dataset. In this analysis, we only consider the predictions that result in a structural match. One observes that the \texttt{DFT} structures exhibit a median energy change during relaxation of only 2~meV/atom. Similarly, the structures from the \texttt{EXP} dataset present a median energy change of 5~meV/atom. This indicates that the generated candidates are already energetically close to their final DFT local minimum (estimated by the MLIP).

To facilitate comparison and analysis with respect to the references, we now also structurally relax the positions of the \texttt{DFT} and \texttt{EXP} references using \texttt{NequIP}, and call the resulting datasets \texttt{DFT+MLIP} and \texttt{EXP+MLIP}, respectively; see Methods section~\ref{ssec:models-and-training}. 
By relaxing the references, indeed, we can directly compare the energies of the prediction (which we always relax with the MLIP in our algorithm, as mentioned earlier, steps 3 and 4) with the initial reference (now relaxed with the same MLIP)
to assess whether the prediction matches the original reference. Moreover, we avoid that different local minima obtained with DFT and MLIPs affect the comparison.
Notably, however, we stress that this is only used for the comparison, but the inpainting process always starts from the original \texttt{DFT} and \texttt{EXP} datasets. 
We highlight that the comparison with the \texttt{EXP} dataset allows us to assess whether we would end up in the same configuration when directly starting from experimental source structures or from our prediction followed by DFT or MLIP relaxations. This does of course not exclude general deviations between experiment and DFT predictions, but is not a limitation of the approach itself, as the inpainting approach can also be applied to training sets computed at higher level of theory, beyond DFT. 
Fig.~\ref{fig:DL:structural-energetic-agreement-MLIP}b then shows the RMSD of the predicted structure with respect to the reference, after full relaxation (here, we normalize by the number of atoms and not the number of hydrogen sites, since all atoms are optimized). The main plot shows that the differences are very small in case of \texttt{+MLIP} datasets, with the majority below 0.01~\AA, indicating that the predictions (after reconstructing the H positions) relax to the same configuration as if one starts from the references (with H atoms) and relaxes them with the same MLIP. 
In SI Fig.~S7
we also compare how close the diffusion model outputs are with respect to the reference datasets without additional relaxation.
Slightly larger differences are observed in this case, as expected due to the typical deviations between the MLIP-relaxed structures and the original ones (DFT-relaxed or experimental). Notably, the \texttt{DFT} case gives an estimate of how much the \texttt{NequIP} MLIP results differ from the DFT-relaxed results (i.e., those in the MC3D database), which is valuable additional information since \texttt{NequIP} is trained on a different training set. 

The observed structural trends for the MLIP-relaxed predictions and targets (''+MLIP'' datasets) are further supported when focusing on the energetic agreement between the predictions and references.
In Fig.~\ref{fig:DL:structural-energetic-agreement-MLIP}c one observes that the cumulative distributions of the energetic differences almost follow a step-like function at around 0~eV when a structural match is observed. This confirms the high structural agreement observed in Fig.~\ref{fig:DL:structural-energetic-agreement-MLIP}b and indicates that, when a match is detected by the \texttt{StructureMatcher}, the prediction essentially always relaxes to the very same configuration.

As expected, and already observed for the structural RMSD, more significant deviations can be observed for the cases that do not result in a structural match. However, and most notably, the majority of those non-matching cases end up in a configuration that is more stable than the structure from the reference source database (energetic difference below 0~eV), always with respect to the potential energy surface described by DFT or the MLIP. Therefore, these should not be considered failures of our approach, but rather as successful predictions that actually identify an even more stable configuration than the initial reference.

To further validate that this result is not an artifact of the MLIP, but that our method truly discovers new more stable configurations, we perform DFT relaxations (only of the positions) and again compare the energetic differences between the inpainted predictions and references. For this analysis, we only consider those structures for which no structural match was obtained for the most stable candidate (114 and 132 for \texttt{DFT} and \texttt{EXP}, respectively, of which 1 and 17 did not converge successfully in DFT and were discarded for the later analysis; see section~\ref{ssec:methods-dft-energy-vali} for further details on the computational details).

\begin{figure}[htb]
\includegraphics[width=\linewidth]{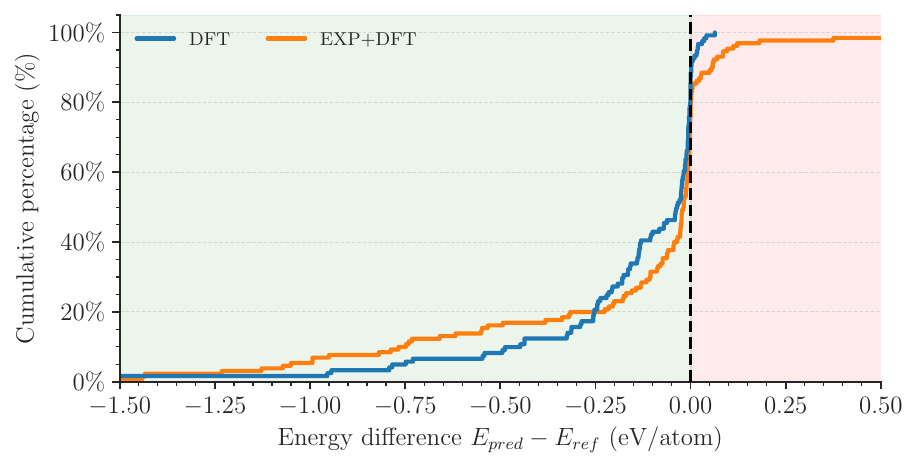}
\caption{
\textbf{DFT validation of the relative stability of predictions and references.}
Cumulative distribution of the energy difference computed with DFT between the original reference and the prediction of the model (the reference and prediction where both optimized by DFT, as indicated by the \texttt{EXP+DFT} in the legend, by relaxing only the atom positions and not the cell). 
The analysis is limited to structures for which the lowest energy sample among 30 samples does not result in a structural match.
For the DFT energy evaluation, we only select the most stable sample according to the MLIP energy evaluation.
The green (red) area indicates predictions that are energetically more (less) stable than the original reference.
}
\label{fig:DL:DFT-energy-validation}
\end{figure}

The cumulative distribution of the energy differences after DFT relaxation is presented in Fig.~\ref{fig:DL:DFT-energy-validation}. 
We obtain that 84.2\% and 77.4\% of the model predictions (for the DFT and EXP datasets, respectively) have an energy that is lower than the original reference (i.e. the predictions are more stable; green area in Fig.~\ref{fig:DL:DFT-energy-validation}), thus validating via DFT what we had observed in the MLIP results: when there is a structural mismatch, our diffusion model typically identifies even more stable structures than the original reference. 
We stress that the aspect of higher stability than the original reference, especially in the context of the \texttt{EXP} dataset, is always with respect to the DFT potential energy surface. Moreover, it always means more stable than the reported configuration in the source databases, which often contain approximated or guessed hydrogen positions. Nonetheless, having in mind the main application outlined in the introduction, i.e., the extension of materials databases based on known host structures, finding energetically more stable configurations (within the framework of a chosen method, e.g., DFT) is indeed one of the main targets of crystal structure prediction applications.

SI section~S1\,C further discusses the correlation of the energy differences predicted by different MLIPs, also justifying the selection of the \texttt{NequIP}~\cite{batzner_e3-equivariant_2022} foundational model~\cite{kavanagh_2025_16980200}, as it is the one---among the five we tested---that best agrees with our DFT settings in predicting the relative stability of the inpainted structures and the references in our setup, and generally belongs to the best performing foundational MLIPs at the time of publication~\cite{riebesell_framework_2025}.
One can rationalize the observation that most of the structurally non-matching predictions are actually more stable since \texttt{MatterGen} (and our extensions) were only trained on materials with an energy above hull smaller than 100~meV/atom, thus naturally including a bias toward predicting stable materials.
Moreover, having in mind the underlying challenge of experimentally determining the hydrogen positions, these apparently ``non-matching'' cases might be those for which the hydrogen positions were only roughly estimated in the experimental source databases.

Based on these insights, we will define a successful prediction in the remaining analyses as either resulting in a structural match, i.e., reconstructing the original reference, or in an energetically more stable configuration\footnote{
The accuracy of these predictions for equilibrium lattice geometries in general, and for bonding interactions of hydrogen in molecules and solids, depends on the approximations used in the simulations, particularly the chosen exchange–correlation functional. While systematic improvements have been achieved with more sophisticated approximations over the past decades, this remains a very active area of research and we refer the interested reader to recent literature on the topic~\cite{Veccham2020,Mitchell2025,Dasgupta2022,Dasgupta2021,Santra2007}.
}. To distinguish these successful predictions from the previous discussion considering only structural matches, we will call our new success rate the \textit{Lower Energy or Structural} (LES) matching rate in the remaining analyses.

\subsection{Maximizing the performance even beyond the training regime}
\label{ssec:results-LES-performance}

\begin{figure}[htb]
\includegraphics[width=\linewidth]{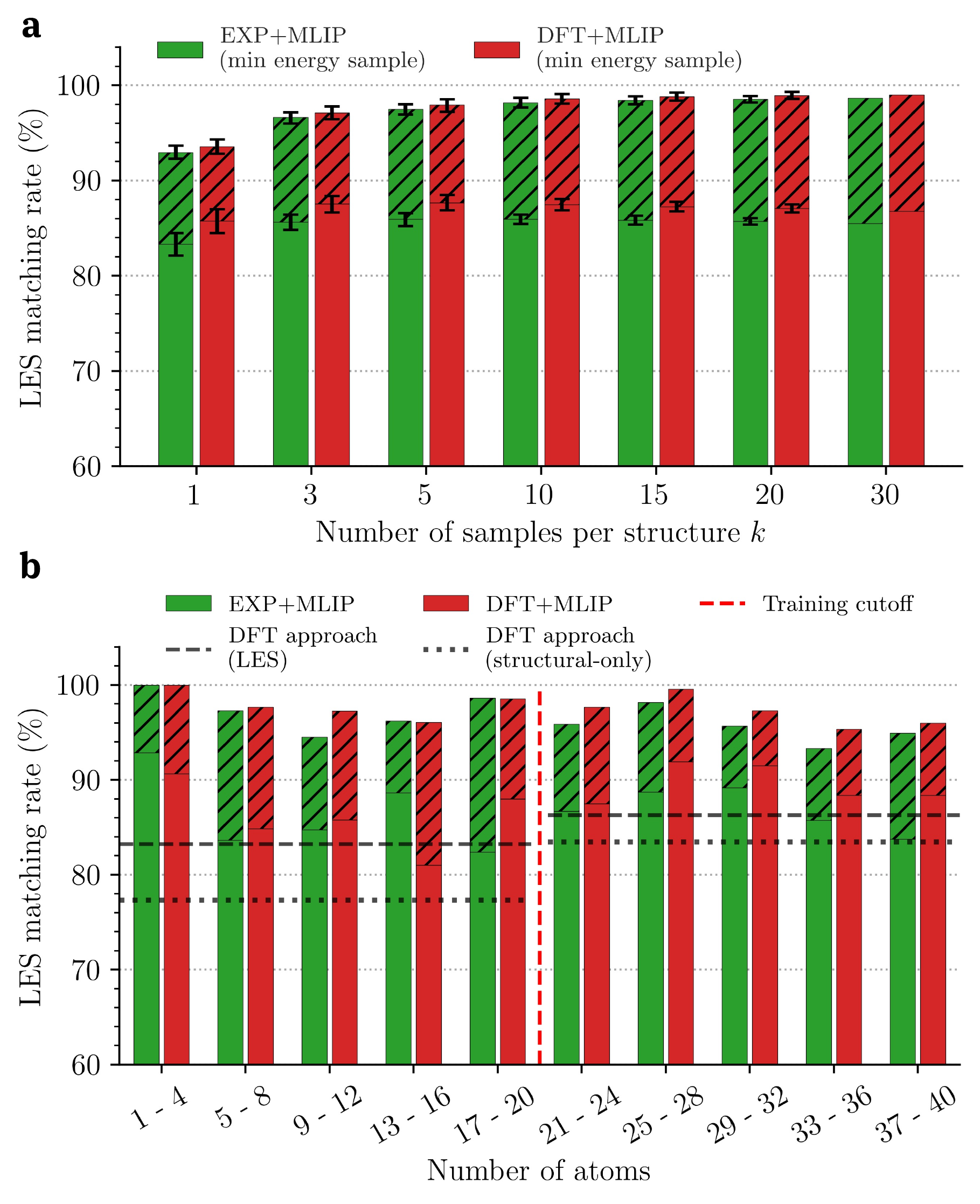}
\caption{
\textbf{Analysis of the multi-trial \textit{Lower Energy or Structural} (LES) matching rate, resolved with respect to the number of samples and unit cell size.}
In both panels, the lower solid portion of each bar shows structural matches, while the hatched upper portion indicates additional successes where the final configuration is not a structural match but is energetically more stable.
\textbf{a} LES matching rate as a function of the number $k$ of generated samples per structure. For each structure, the most stable sample among the $k$ generated samples is selected to compute the LES matching rate. For $k < 30$, uncertainty is estimated via boot-strapping and shown as error bars.
\textbf{b} LES matching rate as a function of the number of atoms per unit cell for the \texttt{EXP+MLIP} and \texttt{DFT+MLIP} datasets, considering only the most stable sample among $k=30$ and $k=10$ samples per structure, for the ones containing up to 20 and 21 to 40 toms, respectively. The vertical red dashed line marks the atom-count cutoff of 20 atoms/cell for the the \mbox{\texttt{Alex-MP-20}} training dataset~\cite{zeni_generative_2025} of MatterGen.
Furthermore, the black dotted and dashed lines indicate the average performance (across the range 1-20 and 21-40 atoms) in terms of structural and LES matches of the direct DFT-based reconstruction approach, discussed in the SI. Those values should only be compared to the red \texttt{DFT+MLIP} dataset.
}
\label{fig:DL:n-atoms-success}
\end{figure}

In Fig.~\ref{fig:DL:n-atoms-success}a we show the newly defined LES matching rate as a function of the number of samples that are generated per structure. We distinguish between pure structural matches (lower part of the bars) and LES matches (upper part of the bars that is ``hatched'', i.e., overlayed with diagonal lines).
The performance on smaller sample sizes $k < 30$ are estimated via bootstrapping, i.e., resampling subsamples with replacement, which is also used to estimate the uncertainty of the success rate. 
We use the energetically most stable configuration among the $k$ samples to determine the structural and LES matches, which is also the recommended algorithm for production runs where no reference is available.
Here, the energy is evaluated with \texttt{NequIP} only due to the large number of samples per structure. A significant performance increase is observed when generating several samples compared to the generation of a single sample.
Overall, we achieve a notable 99\% (98.7\%) LES matching rate for the \texttt{DFT+MLIP} (\texttt{EXP+MLIP}) dataset when generating 30 samples (evaluated with \texttt{NequIP}; therefore, small deviations with respect to DFT might occur, resolved in the next section), highlighting the efficiency of our approach in reconstructing the hydrogen positions given a host lattice. In order to balance performance and computational cost (as a reference, the inpainting process for the 862 structures in the \texttt{DFT} dataset with 300 denoising steps and 1 sample per structure takes approximately 30 mins on an Nvidia A100 GPU), we recommend a value of $k=10$, yielding a performance of 98.6\% (98.1\%) for \texttt{DFT+MLIP} and \texttt{EXP+MLIP}, respectively, with almost no decrease in performance compared to the $k=30$ results. We note that this recommendation applies to high-throughput workflows. Based on the small computational cost, for more targeted applications one should keep $k=30$ samples and potentially even further increase this value.

Another related way to improve the performance, at higher computational costs, is to perform the selection of the energetically most stable sample fully with DFT. As a reminder, here we used \texttt{NequIP} to select the most stable sample and only relax that sample with DFT. Fig.~S8~and~S9 in the SI discuss in more detail the differences in the energetic ranking of the 30 samples for each structure, obtained with DFT and \texttt{NequIP}. In that analysis, we focus on the 18 structures in the \texttt{DFT} dataset that did not result in a LES match. When replacing \texttt{NequIP} with DFT to select the most stable sample, 11 out of the 18 cases also result in a LES match, increasing the LES matching rate to more than 99\%, which is actually in agreement with the MLIP-evaluated result discussed in the context of Fig.~\ref{fig:DL:n-atoms-success}a. This highlights that, even though the MLIP-based energetic ranking of the samples introduces some uncertainties, the overall estimation of the success rate is very reliable. The difference is not necessarily a shortcoming of the \texttt{NequIP} model but could also be related to the different DFT settings (compared to the training set) and can potentially resolved by fine-tuning the foundational model.

Lastly, one relevant aspect is whether the differing configurations occur because of the differing hydrogen positions, or due to a significant change of the positions of the host sites after predicting the hydrogen positions and performing structural relaxations. To quantify this, we used again the StructureMatcher, but this time considering only the non-H atoms, and applied it to compare the MLIP-relaxed prediction (full atomic relaxation at fixed cell, so that the host atoms can in principle move) with the reference (also MLIP relaxed, to be comparable). We observe that only in 4.4\% and 4.1\% of the DFT and EXP structures, respectively, the host positions do not match anymore. Hence, especially in the context of the EXP dataset, one can note that even when the hydrogen predictions do not result in a full structural match, most of the lower energy matches do still match the host structure, which is the part that is assumed to be measured with a higher confidence in experiment. This also hints at the fact that the hydrogen positions and small adjustments of the host sites are the aspects that lead to a lower energy prediction, rather than a major change in the atomic positions of the original host structure.

In addition to the previous improvements, we now discuss how the method yields promising results beyond the training regime. First, as outlined in the Methods section describing the data, we note that we carefully removed any overlapping structures from the training and validation set, with respect to the \texttt{DFT} dataset, when training our \texttt{pos-only} and \texttt{pos-only-TD} models, to ensure a valid comparison. 
One of the current (potential) limitations of the \texttt{MatterGen} model~\cite{lee_generative_2025} (and of our extension) is that it is trained only on structures with up to 20 atoms per unit cell. This is a practical limitation to reduce the computational cost of the training.
While it can be lifted, it is important to verify how well the model performs beyond this limit.
In Fig.~\ref{fig:DL:n-atoms-success}b we therefore show the LES matching rate resolved by the number of atoms per unit cell, up to a value of 40. 
Here, we consider the previous $k=30$ samples for each structure containing up to 20 atoms from the previous analysis, and generate $k=10$ samples for the structures containing 21 to 40 atoms (as this was determined as the optimal trade-off and in order to reduce computational cost). Moreover, as this is used to estimate the final performance, the energies of the most stable sample (selected based on \texttt{NequIP}) were calculated by relaxing the predictions with DFT. Thus, all the LES matches are actually confirmed by DFT (we note that 98 out of 4588 structures for which the DFT relaxation failed were removed).

We observe that the performance of the model remains very large even when applying it to structures much larger than those seen during training. Only for much larger systems ($>32$ atoms per unit cell), a reduction in performance starts to be observed, especially when starting from experimental structures. Nonetheless, the LES matching rate remains well above 90\%. 
In Fig.~\ref{fig:DL:n-atoms-success}c we also indicate the average performance (averaged over systems with 1--20 and 21--40 atoms) for the DFT-based approach that is discussed in detail in the Methods section and the SI. For those results, we also distinguish between pure structural matches and LES matches.
Compared to the DFT-based approach, the diffusion model and MLIP based approach achieves a higher structural matching rate across almost all bins on the \texttt{DFT+MLIP} dataset. Moreover, the \texttt{pos-only-TD} predictions also shows a higher ratio of lower energy matches in addition to the purely structural matches.

In summary, this analysis highlights that the method that we present here works well even beyond the training regime, although with some expected degradation in the performance. This is especially promising as our model can act as a foundation to be efficiently fine-tuned with larger structures to further increase its performance, without the necessity of full retraining. However, this goes beyond the scope of the present work and is left for future work.

\begin{table}[h]
    \centering
    \caption{
    \textbf{Summary of the final algorithm.} Performance of our final algorithm based on the \texttt{pos-only-TD} model and $k$ samples per structure. The performance is shown for the \texttt{DFT+MLIP} and \texttt{EXP+MLIP} dataset containing up to 20 atoms (i.e., the training regime). Furthermore, the performance on structures with 21 to 40 atoms per unit cell is shown in brackets, as a reference for the performance beyond the training regime.
    }
        \begin{tabular}{ccccc}
              \phantom{x}model\phantom{x} & \phantom{x}$k$ samples \phantom{x} & \phantom{x} DFT+MLIP\phantom{x} & \phantom{x} EXP+MLIP \phantom{x} \\
              \colrule
             \texttt{pos-only-TD} & 10 & 97.8\% (97.2\%) & 97.2\% (95.6\%) \\
        \end{tabular}
    \label{tab:DL:n-atoms-success}
\end{table}

\begin{figure*}

\includegraphics[width=\linewidth]{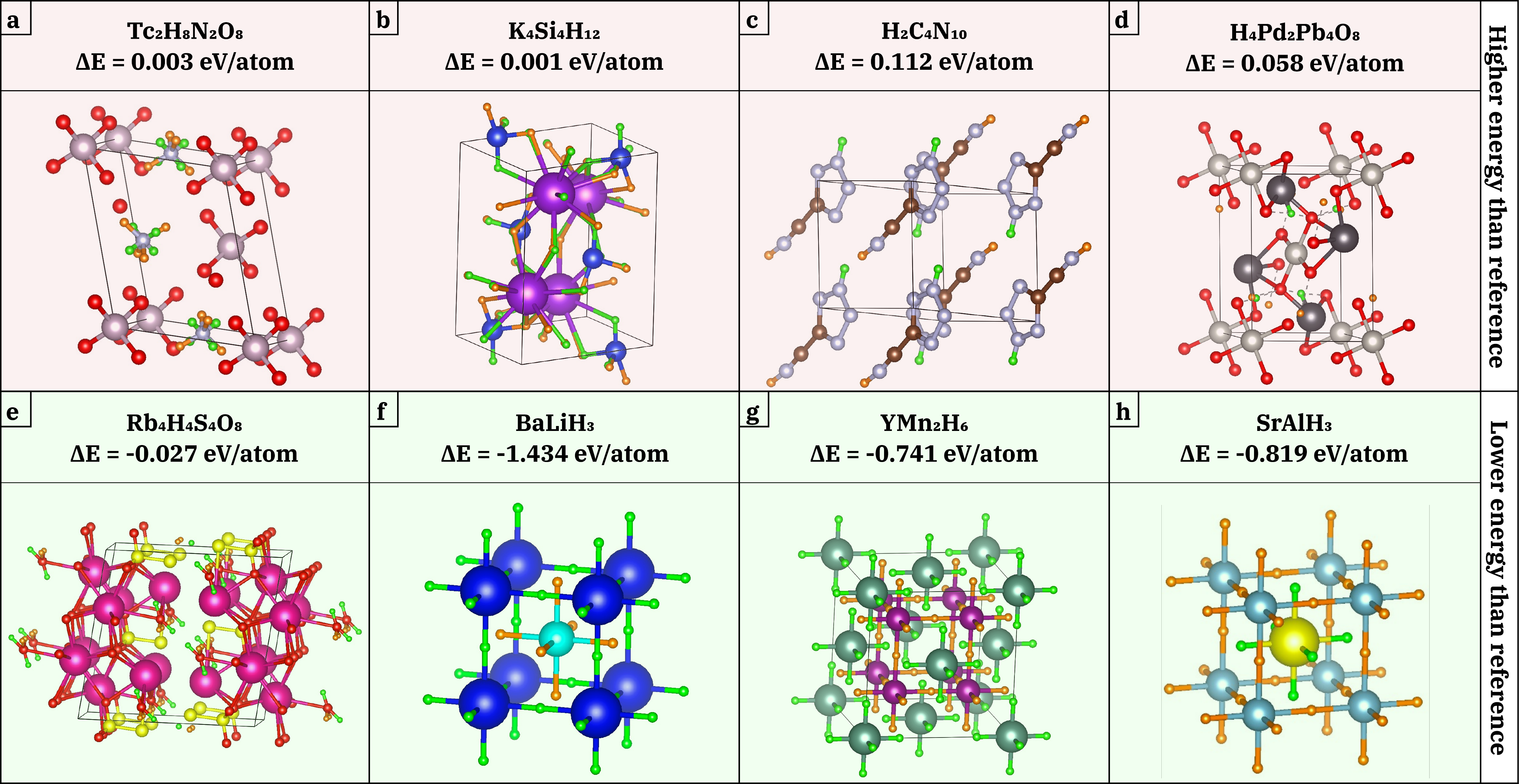}
\caption{
\textbf{Visualization of predicted hydrogen positions that do not result in a structural match.} The examples are taken from the \texttt{EXP} dataset.
In the panels, the small green and orange spheres represent the positions of hydrogen atoms in the reference structures and in our approach, respectively.
\textbf{a-d} Four example structures that do not result in a structural match and exhibit a higher energy than the experimental reference structure (evaluated after a subsequent DFT relaxation). The IDs in the source databases are: a)~ICSD-10429, b)~MPDS-S1706693, c)~MPDS-S1129483, d)~ICSD-195137.
\textbf{e-h} Examples that do not result in a structural match but relax to configurations that are lower in energy than the original references.
The IDs in the source databases are: e)~ICSD-425888, f)~MPDS-S1903387, g)~MPDS-S1832431, h)~ICSD-291483. 
}
\label{fig:DL:failure-more-stable-examples}
\end{figure*}

Based on the previous discussion, we summarize our recommended algorithm in Table~\ref{tab:DL:n-atoms-success}, together with the corresponding performance. We use our newly trained \texttt{pos-only-TD} model, i.e. the \texttt{MatterGen} architecture in combination with the \texttt{TD} approach~\cite{mayet_td-paint_2024} to only denoise positions, \mbox{generate~$k=10$} samples per structure, and select the lowest energy one as the final prediction (either based on DFT or MLIPs, depending on the trade-off between computational cost and slightly higher accuracy; as outlined before, using DFT for all samples could even further boost the performance).


\subsection{Inspecting mismatches and lower energy matches}
\label{ssec:explicit-examples}
After having discussed the generally very high LES matching rate, we now explicitly discuss in more detail some of the mismatching predictions of our algorithm (i.e., cases in which our algorithm predicts a higher-energy structure than the reference) and some of the predictions that are more stable than the original reference. We focus on the \texttt{EXP} dataset for this analysis, but the observed patterns are similar also in the case of the \texttt{DFT} dataset.

We first note that several of the mismatches actually correspond to structures that are labeled as theoretical in the MC3D database.
This label is derived from the information available in the underlying source databases and typically indicates structures that have been just postulated and then used in simulations, but never experimentally identified.
If we filter out such theoretical structures, thus reducing the \texttt{EXP} dataset to 577 structures, we only find five structures for which the prediction does not result in a LES match. 
Hence, when only focusing on structures that are actually experimentally known, the success rate even exceeds 99\%. 

Fig.~\ref{fig:DL:failure-more-stable-examples}a-d present four examples that do not match the reference and result in a configuration that is higher in energy. We note that the visualized crystal structures show the prediction after the constrained MLIP relaxation, but the $\Delta E = E_{inpainted} - E_{ref}$ values correspond to the energy differences obtained after a subsequent DFT relaxation (of the reference and prediction) as explained earlier.

In the cases of Tc$_2$H$_8$N$_2$O$_8$ and K$_4$Si$_4$H$_{12}$ (Fig.~\ref{fig:DL:failure-more-stable-examples}a, b), the final energy difference is very small and is only based on slightly different orientations of the hydrogen positions, with some of them actually overlapping with those in the original reference. Moreover, Tc$_2$H$_8$N$_2$O$_8$ (and also the fifth structure that does not result in a LES match, K$_4$H$_8$Pt$_2$, not shown here), belong to the examples for which a different sample identified by the algorithm would actually match the reference, but the lowest energy one according to \texttt{NequIP} does not. 
In the case of the molecular crystal H$_2$C$_4$N$_{10}$ (Fig.~\ref{fig:DL:failure-more-stable-examples}c), the hydrogen site is essentially located at the opposite position of the molecule and results in a significantly higher energy.
Similarly, two of the four hydrogen sites in H$_4$Pd$_2$Pb$_4$O$_8$ (Fig.~\ref{fig:DL:failure-more-stable-examples}d) are located close to their reference positions, while the other two are located differently and also show a different bonding behavior, and result in an energy that is 58~meV/atom higher than the energy of the reference. 

We now discuss the examples in Fig.~\ref{fig:DL:failure-more-stable-examples}e-h, for which the predictions are energetically more favorable than the reference. 
The underlying reference~\cite{Rb4H4S4O4-example-Lehner2013} of Rb$_4$H$_4$S$_4$O$_8$ (Fig.~\ref{fig:DL:failure-more-stable-examples}e) reported that the unambiguous determination of the hydrogen positions of the water molecules was not successful (the measurement was based on XRD). The water molecules are slightly rotated in our prediction to be aligned within the layers, and the relaxed configuration is more stable according to DFT.
The examples of BaLiH$_3$ and YMn$_2$H$_6$ (Fig.~\ref{fig:DL:failure-more-stable-examples}f, g) correspond to cases where it seems that issues might have occurred in the transcription from the experimental papers to the source databases. Our predictions do not match the references obtained from the source databases, but they actually agree with the experimental data from the original papers. These examples highlight how our method can help to potentially identify insufficient data quality and detect structures that might require more careful inspection.
Finally, SrAlH$_3$ (Fig.~\ref{fig:DL:failure-more-stable-examples}h) is an example in which our method can actually help in the refinement process. The underlying paper already stated that the determination of the hydrogen sites using XRD was difficult due to the low X-ray scattering~\cite{Kamegawa2014}. Our prediction indicates that the assignment of the hydrogen sites to a different Wyckoff position in the perovskite structure results in an energetically much more favorable configuration (while still having a very similar XRD pattern, see SI Fig.~S10 for a comparison).

More details on these examples can be found in SI section~S1\,F. Furthermore, we point the interested reader to SI section~S1\,G for further comments about the application of our method to experimental structures to support the refinement process.

\section{Discussion}
\label{sec:discussion}

In summary, we presented how inpainting techniques, originally developed in the field of image vision, can be adapted in materials science to accurately and efficiently reconstruct crystal structures from partially known information. In particular, we extended the recently published \texttt{MatterGen}~\cite{zeni_generative_2025} model adapting it for crystal structure inpainting, by leveraging the \texttt{TD-Paint} inpainting technique from computer vision and applied it to the problem of finding missing hydrogen positions in inorganic crystal structures. The method has been successfully benchmarked on a set of reference structures with known hydrogen positions (obtained from three different experimental crystallographic databases), by artificially removing hydrogens and then reconstructing their positions using our method. The results show that the inpainting approach significantly outperforms both unconditioned diffusion models and previous DFT-based approaches,
leading to an overall success rate of~97\% (defined as either structural matches with the reference structures, or an even more energetically stable prediction). 
In case of the \texttt{EXP} dataset, this success rate can be pushed even further to~99\% when excluding structures flagged as theoretical in MC3D. Moreover, the same success rate can be achieved for the \texttt{DFT} dataset by replacing MLIP energy evaluations with DFT calculations to select the most stable candidate among those identified by the algorithm.
We also show how the model performs beyond the training regime, achieving high performance on structures containing up to 40 atoms per unit cell, although being trained only on structures with up to 20 atoms per unit cell. Moreover, by explicitly discussing some of the mismatches as well as predictions that are energetically more stable than the original references, we also highlight how the method can help to detect cases for which the underlying experimentally derived data might require additional checks.

Our method and trained models are completely hydrogen-agnostic and can thus also be directly applied to any other problem where one seeks to place additional sites into a host crystal structure. One example application is the study of Li intercalation in cathode materials to identify the most stable Li arrangement for partially lithiated states, and similarly for other ions (such as Na in sodium-battery materials). In such applications, it might be of interest to re-enable the adjustment of the unit cell during the denoising process, which was not required for the present study, but might be beneficial when intercalating ions into pristine materials.
Another example is the prediction of potential muon stopping sites, which normally requires the generation of several symmetrically distinct trial arrangements~\cite{moller_quantum_2013,moller_playing_2013,onuorah_automated_2025}. By leveraging the method presented in this work, this process could potentially be accelerated by directly generating fewer promising trial configurations. Furthermore, hydrogen-related research topics might benefit from this work for finding interesting candidate structures that could potentially be studied in more detail. The main applications we envision are two-fold. First, extending computational databases (such as, e.g., MC3D)  with several additional hydrogen-containing materials will enable computational screening studies that might detect hydrogen-containing materials with desired properties that had not been considered before. Second, similar to the aforementioned example of partially lithiated cathode materials, one could screen host materials and search for favorable positions of interstitial hydrogen sites.

Finally, we leave the determination of the number of (missing) sites to be inpainted for future work, as the main goal here was to demonstrate the high performance of the method. In the context of missing hydrogens, this is also motivated by the fact that experimental crystallographic files often report the expected number of missing hydrogen sites, e.g. based on other experimental techniques or chemical reasoning. Moreover, in other applications, such as the intercalation mentioned before, the number of sites to be added is anyway known.
A potential extension towards the determination of the number of missing sites could be to screen a range of sites to be added $N_{inpainted} \in \{1, \dots, N_{max}\}$ and exclude unstable ones based on a convex hull and formation energies approach. While this will still yield several possible $N_{inpainted}$, it already presents a first step to exclude unlikely compositions. A promising proof of concept on a small set of structures is discussed in the SI section~S1\,B.

All the methods and extensions on top of \texttt{MatterGen} presented in this work are publicly available in our \texttt{XtalPaint} package~\cite{web:xtalpaint} and also integrated into the \texttt{AiiDA} framework~\cite{pizzi_aiida_2016,huber_aiida_2020, uhrin_workflows_2021}.

\section{Methods}
\label{sec:method}

\subsection{Score based diffusion models}

In this work, we use \texttt{MatterGen}~\cite{zeni_generative_2025} as the underlying diffusion model. Except for small technical modifications which are described in the following two sections, the model architecture remains the same. 
Nonetheless, hereafter, we briefly summarize the diffusion and denoising of the fractional coordinates. The original \texttt{MatterGen} was trained to also denoise the cell and atomic types. However, this is not used in the present study as removing this degree of freedom improves the model result for our application, as we show in Fig.~\ref{fig:DL:compare-inpainting-methods}a. We refer the interested reader for further details on the architecture and original training to the original reference~\cite{zeni_generative_2025}. 
As shown by Song \textit{et al.}~\cite{song_score-based_2021}, score-based diffusion models can be formulated as forward and reverse processes governed by a stochastic differential equation (SDE). The forward process is given by 
\begin{equation}
    dx = f(x, t)dt + g(t)dw,
\end{equation}
where $f(x, t)$ is the drift coefficient, $g(t)$ is the diffusion coefficient, and $dw$ represents standard Brownian motion. The corresponding reverse process, transforming noise into actual samples, is governed by
\begin{equation}
dx = \left[ f(x, t) - g^2(t)\nabla_x \log p_t(x) \right] dt + g(t) dw.
\end{equation}
Here, $\nabla_x \log p_t(x)$ is the score function approximated by our score network.
In particular, the forward process is following a so-called Variance-exploding SDE (VE SDE)
\begin{equation}
dx = \sqrt{\frac{d[\sigma^2(t)]}{dt}} dw,
\end{equation}
where $\sigma(t)$ is an exponentially increasing sequence of standard deviations.

In practice, the process is discretized and numerically iterated, as described in the following~\cite{zeni_generative_2025}:
\begin{equation}
x_{t-1} = x_t + \left( \sigma_t^2 - \sigma_{t-1}^2 \right) s_{\theta^*}(x_t, t) 
+ z \sqrt{\frac{\sigma_{t-1}^2 \left( \sigma_t^2 - \sigma_{t-1}^2\right)}{\sigma_t^2}},
\label{eq:reverse-discretized}
\end{equation}
where $z \sim \mathcal{N}(0, I)$ is sampled from a standard Gaussian, $\sigma_t$ is the noise level at timestep $t$ and $s_{\theta^*}(x_t, t)$ is the score function approximated by our score network. These results show the process for an unconditional model. However, for our inpainting task, we do not want to adjust all positions but rather mask certain sites in the structure. The next subsection discusses this aspect in more detail.

\subsection{Improved inpainting techniques}
\label{ssec:methods-inpainting}
While unconditioned diffusion models can be directly used for inpainting tasks~\cite{song_score-based_2021}, i.e., completing and reconstructing a masked region in an image, several approaches have been proposed in the field of computer vision to improve their performance~\cite{lugmayr_repaint_2022,zhuang_task_2024,corneanu_latentpaint_2024}. Among others, Luan \textit{et al.} proposed the so-called \texttt{RePaint} approach~\cite{lugmayr_repaint_2022} which has also been adopted in other fields, such as materials science and computational chemistry~\cite{Zhong2025, schneuing_structure-based_2024}. The motivation was the observation that in images, but also molecules, the inpainted regions are often not fully homogenized with respect to the known region. 

The \texttt{RePaint} approach addresses this inconsistency by going back and forth in the denoising process, by adding corresponding noise level at timestep $t_i$ to the known region. Since the noise on the known regions is not passed between iterations, i.e., it is always added to the exactly known values, the final prediction will still contain the correct known part. In addition to the pure number of resampling steps, one can also specify the so-called jump length that defines for how many steps the noise is added again, before continuing the denoising process for the same number of steps. 
Although \texttt{RePaint} can significantly improve the performance, it comes at the cost of potentially many more denoising steps. Fig.~S3 in the SI shows the denoising schedule for some parameter combinations that have been used in other works~\cite{Zhong2025}, i.e., 200 noise levels (``standard'' denoising steps), 3 resample iterations and a jump length of 10. This results in 961 steps compared to the 200 one that one would use when just applying the unconditional diffusion model. 
In the context of crystal structures, one can rationalize this in the following way. The model is trained under the assumption that all the sites exhibit the same level of noise, as one timestep $t_i$ is sampled per structure during training. While this is true for actual structure generation tasks, in the context of inpainting the known region should essentially not contain any noise and only the inpainted region should be updated. 

To avoid this significant increase in the number of steps that makes the generation process much slower, further approaches have been proposed. One example is \texttt{TD-Paint}~\cite{mayet_td-paint_2024}, which also addresses the problem of the aforementioned inconsistency. 
Transferring the underlying idea from computer vision to the present problem of inpainting missing sites in crystal structures, this approach retrains the model so that it ``understands'' that different sites in a given structure can have different noise levels, so that it can always condition on the known information without adding noise to it. In practice, in the context of inpainting, we first randomly sample one timestep $t_i$ per structure, that is identical for all sites, as done before. Afterwards, we replace the timestep for $p$ percent of the sites in a training batch with a timestep $t=\epsilon$ that is close to 0, which corresponds to a small level of noise. The presented model uses a fraction of $p=20\%$. In this way, the model is aware of the fact that some sites in a structure contain a given level of noise, while others are essentially noise-free and do not have to be updated. The actual denoising step according to Eq.~\eqref{eq:reverse-discretized} can then be performed as before, with the small adjustment of combining the known and inpainted sites: 
\begin{equation}
x_{\tau} = x_{t}^{\ominus} \odot (1 - m) + x_{0}^{\oplus} \odot m,
\end{equation}
where $\tau$ is the adjusted time consisting of the noisy timestep $t_i$ and the $t_{\epsilon}$ for the inpainted region.

\subsection{Model training and data}
\label{ssec:models-and-training}
Within the main text, we refer to four different models that are described in this section. As mentioned before, we leverage the \texttt{MatterGen} architecture (more details in~\cite{zeni_generative_2025}) as the underlying diffusion model in this work and modify it to handle individual timesteps per site. We also keep the original training and sampling parameters, only adjusting the signal-to-noise ratio in the Langevin dynamics of the sampling process from 0.4 to 0.2 (see SI Fig.~S11).
We refer the interested reader for further details on the curation of the training set and the underlying model architecture to the corresponding reference~\cite{zeni_generative_2025}. The different models and their differences are briefly listed below:
\begin{enumerate}
    \item \texttt{MatterGen}: the baseline model from Ref.~\cite{zeni_generative_2025}, originally trained to denoise the positions, lattice and atomic types, even though the lattice and atomic types are fixed for the denoising in our inpainting application.
    \item \texttt{pos-only}: a newly trained model to only denoise the positions.
    \item \texttt{pos-only-RePaint}: the previous newly trained \texttt{pos-only} model, but evaluated in combination with the \texttt{RePaint} algorithm~\cite{lugmayr_repaint_2022}.
    \item \texttt{pos-only-TD}:  newly trained with different noise levels per structure, following the concept of \texttt{TD-Paint}~\cite{mayet_td-paint_2024}. In this case, we adjust and extend the \texttt{MatterGen} source code to be compatible with the definition of noise levels per site instead of a single noise level per structure, which is then expanded for each site in a structure. 
\end{enumerate}
To showcase the performance of our approach, we test the success rate of reconstructing unknown hydrogen positions on the structures containing hydrogen from the MC3D database~\cite{huber_mc3d_2025}. The MC3D database was chosen as it is focusing on, and starting from, experimentally known compounds. Therefore, the issue of unknown hydrogen positions is of high practical relevance in order to extend the coverage of the database. We distinguish two groups of datasets throughout this work:

\begin{enumerate}
    \item \texttt{EXP} and \texttt{DFT}: These are the initial experimental structures that form the foundation of MC3D and their DFT-relaxed counterparts, respectively. EXP contains 915 and DFT 862 structures, respectively. The \texttt{EXP} dataset is slightly larger as it also contains the structures for which the DFT calculations failed when constructing MC3D.
    \item \texttt{DFT+MLIP} and \texttt{EXP+MLIP} The same datasets as before, but where each structure is further relaxed using the \texttt{NequIP} MLIP (only the positions, the unit cell is kept unchanged throughout this work). These datasets are used in the manuscript to assess whether our approach is generally suitable to predict the correct starting point that would relax to the same configuration (even if the outcome of the diffusion model might differ from the original reference). This aspect is particularly important for the construction of DFT curated databases, where one is mainly interested in getting the same configuration after structural optimization.
\end{enumerate}

In section~\ref{ssec:results-LES-performance}, we also extend the discussion to unit-cell sizes beyond the training regime, by considering structures with up to 40 atoms per unit-cell. This extension of the \texttt{DFT} and \texttt{EXP} datasets (21 to 40 atoms) contain 1276 and 1709 structures, respectively (again, we remove any compounds that we could not structurally match with our approach nor validate with DFT).

To ensure a robust benchmarking, we slightly adjust the \texttt{MatterGen} training set, \texttt{Alex-MP-20}, when retraining the \texttt{pos-only} and \texttt{pos-only-TD} models.
Any structure which resulted in a match with a structure in the \texttt{DFT} and \texttt{EXP} datasets (according to \texttt{pymatgen}'s \texttt{StructureMatcher}) was removed when retraining the \texttt{pos-only} and \texttt{pos-only-TD} model. In a post-processing step, we also identified the prototypes (\texttt{aflow\_sym\_label:chemsys}, where  \texttt{aflow\_sym\_label} refers to the AFLOW label~\cite{goodall_rapid_2022,parackal_identifying_2024} and \texttt{chemsys} to the alphabetically sorted chemical system) of each structure and removed any structure from the evaluation that resulted in a matching prototype, to further reduce the risk of any ``data leakage''.  After these filtering steps, the dataset sizes reduce to 831 and 776 in case of \texttt{DFT} and \texttt{EXP}, respectively (the datasets with 21 to 40 atoms do not exhibit any overlap). The similarity of the training and testing datasets is further discussed in the SI Fig.~S12.

\subsection{DFT calculations to validate stability}
\label{ssec:methods-dft-energy-vali}
All the DFT calculations to validate the energetic stability of the references and our inpainted predictions were performed using \texttt{Quantum ESPRESSO}~\cite{Giannozzi_2009,Giannozzi_2017} in combination with the corresponding AiiDA plugin, \texttt{aiida-quantumespresso}~\cite{web:aiida-quantumespresso}, the latest computational protocols~\cite{nascimento_accurate_2025}, and the pseudopotentials from the \texttt{SSSP~PBE~Efficiency~v1.3.0} library~\cite{prandini_precision_2018}. The computational parameters are discussed in detail in the aforementioned references. Nonetheless, we list the main parameters in the following: the k-point mesh is chosen to ensure a maximum distance of at least 0.15~\AA{}$^{-1}$ in each direction in reciprocal space. The force and energy thresholds for ionic optimization are set to $10^{-4}$~Ry/bohr and $10^{-5}$~Ry/atom, respectively. The wavefunction and charge density cutoffs are chosen based on the recommended cutoffs provided by the \texttt{SSSP~PBE~Efficiency v1.3.0} (\url{sssp.materialscloud.org}).

The same DFT parameters were used for the DFT based reconstruction approach, with the only two exceptions that we used the scalar-relativistic pseudopotentials from PseudoDojo~\cite{vanSetten2018} version 0.4 with PBE as the exchange correlation functional, and a cold smearing which is set to 0.01 Ry instead of 0.02 Ry.

\subsubsection{Structural matches}
Structural matches were determined using \texttt{pymatgen}'s \texttt{StructureMatcher} with the default tolerances \texttt{ltol=0.2, stol=0.3 and angle\_tol=5}. Additional checks were done and decreasing the thresholds had only minor impact on the matching rate. Also, compared to other works related to crystal structure prediction, the thresholds adopted here are sometimes even a bit tighter (which is intended, as we only predict parts of the structure and not the whole structure).

\subsection{DFT based algorithm}
\label{ssec:DFT-reconstruction}

We also compare our diffusion-model based strategy against a more physically motivated approach based on DFT that relies on the electrostatic potential to identify candidate hydrogen sites.
Additional electrons, equal to the number of absent H atoms, are introduced and the resulting electrostatic potential
\begin{equation}
V(\mathbf r)=\sum_A \frac{Z_A}{|\mathbf R_A-\mathbf r|} - \int \frac{\rho'(\mathbf r')\, d\mathbf r'}{|\mathbf r'-\mathbf r|}
\end{equation}
is evaluated to identify electron-rich regions. Local maxima of $V(\mathbf r)$, sampled on a fine grid and filtered by distance and proximity criteria, are taken as candidate H sites. When the number of maxima matches the missing H atoms, hydrogen sites are placed at these sites and relaxed, while keeping the host atoms fixed. If the number of candidate sites exceeds the required H atoms, a combinatorial ``pinball'' scheme is applied~\cite{kahle_modeling_2018}, in order to find the energetically most favorable one among all possible arrangements. The process is iterated in a greedy manner until all hydrogen sites are restored, followed by a full structural relaxation.

The algorithm is implemented in Python within the AiiDA infrastructure as the \texttt{RestoreHydrogenWorkChain} workflow~\cite{web:aiida-hydrogen-restorer}, interfacing with Quantum ESPRESSO for self-consistent DFT runs and using peak-finding routines from \texttt{skimage}~\cite{van2014skimage}.
More details on the DFT based approach in SI section~S2, in particular Fig.~S13~and~S14 in terms of the performance and Fig.~S15 related to the challenges of this approach.

\section{Data availability}
All data generated in this work, as well as scripts to generate relevant plots, are available on the Materials Cloud Archive~\cite{Talirz2020} at \url{https://doi.org/10.24435/materialscloud:gz-mt}~\cite{mc_entry}. This entry also includes AiiDA~\cite{huber_aiida_2020} archive files with the full provenance of all DFT simulations, ML calculations and data.

\section{Code availability}
The \texttt{XtalPaint} code~\cite{web:xtalpaint} is publicly available on GitHub: \url{https://github.com/psi-lms/XtalPaint}.
The code is based on the most recent workflow developments of \texttt{aiida-workgraph}~\cite{web:aiida-workgraph} and \texttt{aiida-pythonjob}~\cite{web:aiida-pythonjob}, so that it is possible to use the whole code also without AiiDA~\cite{pizzi_aiida_2016,huber_aiida_2020,uhrin_workflows_2021} as pure \texttt{Python} functions. However, if the user wants to take advantage of remote execution and provenance tracking, one can simply take advantage of the corresponding AiiDA counterparts in the repository.

\section{Acknowledgements}
This research was supported by the NCCR MARVEL, a National Centre of Competence in Research, funded by the Swiss National Science Foundation (grant number 205602).
This work was supported by a MARVEL INSPIRE Potentials Master's Fellowship.
We acknowledge access to Alps at the Swiss National Supercomputing Centre, Switzerland under MARVEL's share with the project ID mr32. Moreover, we acknowledge the use of the Merlin7 cluster run by the Paul Scherrer Institute PSI. M.B. and G.P. acknowledge financial support by the SwissTwins project, funded by the Swiss State Secretariat for Education, Research and Innovation (SERI).
We acknowledge fruitful discussions with  Leonid Kahle, Aris Marcolongo, Nicola Marzari and Daniele Pontiroli.

\section*{Author contributions}
G.P. and P.B. conceived the project. T.R. contributed to the planning of the project, implemented the \texttt{XtalPaint} code, and performed the calculations and analyses related to the score-based diffusion model approach, under the supervision of G.P.
A.C. implemented the DFT based approach and performed all the related calculations and analyses, under the supervision of M.B., P.B. and G.P. 
T.R. wrote the initial version of the manuscript with input from A.C., M.B., P.B. and G.P. All authors discussed the results and contributed to editing and the final version of the manuscript.

\section*{Competing interests}
The authors declare no competing financial or non-financial interests.

\bibliographystyle{apsrev4-1}
\bibliography{refs}

\clearpage
\onecolumngrid



\section*{Supplementary material}

## S1. SUPPLEMENTARY INFORMATION – DEEP LEARNING BASED APPROACH

### A. Extended method comparison

#### 1. Impact of MLIP relaxation on matching rate

Fig. S1 extends the comparison of the different inpainting approaches (see Fig. 1a in the main text) by also showing the structural matching rate without the constrained relaxation of the hydrogen sites (the relaxed results are shown on the right hand side of Fig. S1 and also in the main text). One clearly sees that the relaxed performance is generally better than the direct output of the diffusion model, indicating that the output is not yet in the configuration corresponding to the local minimum. Nonetheless, this also indicates that in these cases the output of the diffusion model is still reasonably close to the final (local) minimum. The strongest difference and improvement is observed for the (original) `MatterGen` model. Already the `pos-only` model performs much better, even without relaxation. Furthermore, the `pos-only-TD` model shows essentially the same performance, only the variance is slightly larger when directly considering the output of the diffusion model. Once again, this highlights the ability of the `pos-only-TD` model to efficiently condition the inpainted sites on the known information. The error bars indicate the minimum and maximum structural matching rate that was obtained among 4 individual runs.

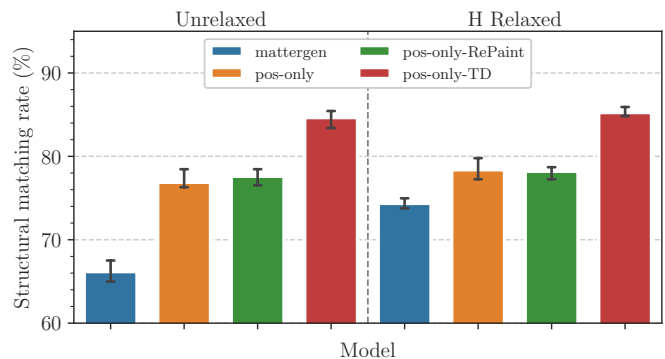

SUPPLEMENTARY FIG. S1. **Impact of (constrained) relaxation on the structural matching rate for different inpainting approaches.** Structural matching rate for multiple models and inpainting approaches: `MatterGen`, `pos-only`, `pos-only-RePaint` and the `pos-only-TD`. Left: structural matching rate for the unrelaxed structures, i.e., the output of the diffusion model. Right: structural matching rate after performing a constrained structural relaxation, i.e., only the positions of the hydrogen sites are optimized, using `NequIP`.

#### 2. TD-Paint vs. RePaint

Fig. S2 further compares different parameter setups for the models using the `TD-Paint` and `RePaint` approaches. The main point to highlight is that while the `RePaint` models do indeed improve the matching rate compared to the `pos-only` model, a lot of this improvement seems to be related to the larger number of denoising steps. This is also true for the impact of the number of corrector steps in the predictor-corrector sampling. The `pos-only-TD` models indicate the number of corrector steps $N_{corr}$ and denoising steps $N_{steps}$ in the labels `pos-only-TD-`$N_{corr}$`-`$N_{steps}$. Moreover, the `RePaint` models follow the following convention: `pos-only-TD-`$N_{corr}$`-`$N_{steps}$`-`$N_{jump}$`-`$N_{res}$, where $N_{jump}$ is the so-called jump length and $N_{res}$ the number of resampling steps (see Ref. [41]). First of all, one observes that the number of corrector steps does not have a significant impact except for the increased number of steps, e.g., comparing `pos-only-TD-5-300` with `pos-only-TD-1-1000`. Similarly, the significant performance gain in case of the `pos-only-RePaint` models is only observed when significantly increasing the number of steps. As an example, `pos-only-RePaint-1-200-3-10` does not really improve compared to `pos-only-RePaint-1-300`, except for a reduction in the spread. While `pos-only-RePaint-1-200-10-10` improves over our standard `pos-only-1-300`, it significantly increases the number of steps, see section S1 A 3. Again, the number of steps (noise levels) seems to be more relevant than increasing the number of corrector steps, as shown for `pos-only-RePaint-1-200-10-10` and `pos-only-RePaint-5-200-3-10`.

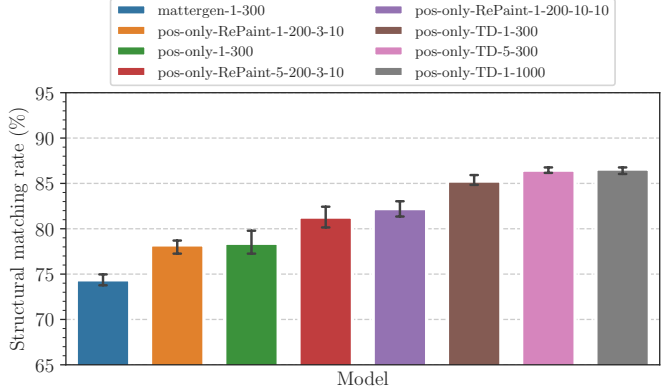

SUPPLEMENTARY FIG. S2. **Comparison of several `TD-Paint` and `RePaint` based models and the impact of different choices of parameters.** The following naming conventions are used: `pos-only-TD-`$N_{corr}$`-`$N_{steps}$ and `pos-only-TD-`$N_{corr}$`-`$N_{steps}$`-`$N_{jump}$`-`$N_{res}$, where $N_{corr}$ is the number of corrector steps, $N_{steps}$ the number of denoising steps, $N_{jump}$ is the so-called jump-length and $N_{res}$ the number of resampling steps (see Ref. [41]).

### 3. Computational cost of the RePaint approach

As already mentioned in the main text and the previous section, while the `RePaint` approach is commonly adopted and improving the performance, it can significantly increase the number of denoising steps and thus the computational costs. In Fig. S3, two common choices [41, 50] of the number of resampling steps and the so-called jump length (see Ref. [41]) are compared, and highlight the significant increase in the number of steps that are performed.

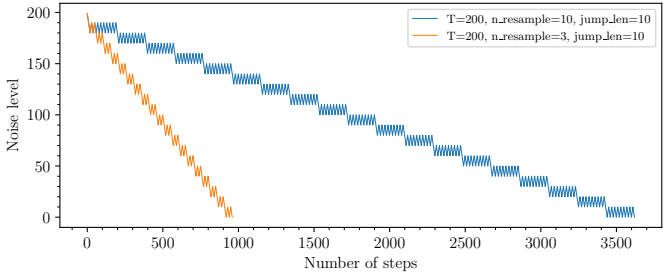

SUPPLEMENTARY FIG. S3. **Increasing number of steps in the `RePaint` approach.** The timestep (corresponding to a certain noise level) as a function of the actual number of denoising steps that are performed for two different setups of the `RePaint` [41] algorithm. Without the `RePaint` approach, the number of timesteps would be equal to the total number of timesteps.

### 4. Impact of the SNR parameter on the performance

The impact of the signal to noise ratio (SNR)—which is used in the Langevin dynamics process and generally impacts how diverse the generated results are—on the structural matching rate is visualized in Fig. S4. For this analysis, we only use the `pos-only-TD` model. One observes that the performance only slightly varies across different SNR values and remains almost constant between 0.1 and 0.6. Nonetheless, it identifies an optimal value of $SNR = 0.2$ (that was chosen in the main text) as it exhibits both the highest performance and smallest spread of the minimum and maximum performance across the 4 independent runs.

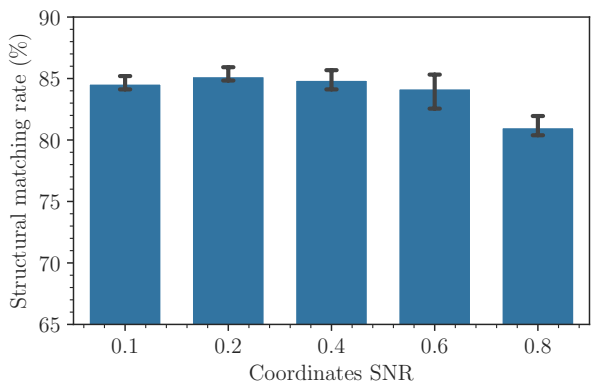

SUPPLEMENTARY FIG. S4. **Impact of the SNR on the structural matching rate.** Structural matching rate as a function of the SNR that is used in the Langevin dynamics during the reverse diffusion process.

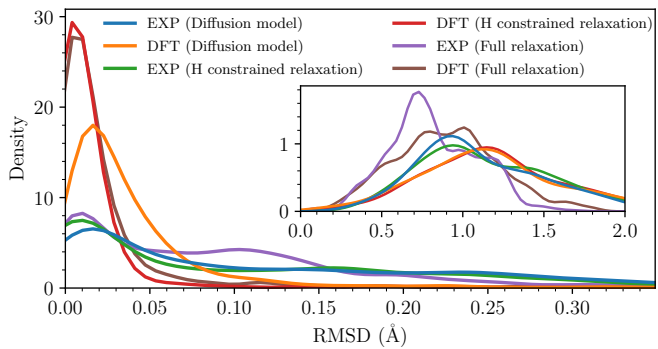

SUPPLEMENTARY FIG. S5. **Impact of different relaxation approaches on the structural agreement of the predictions.** Distribution of the RMSD between the predictions and `DFT` and `EXP` reference. We distinguish two versions of each dataset: On the one hand, the output directly taken from the diffusion model, indicated by the suffix "(Diffusion model)". On the other hand, the prediction of the diffusion model after performing a constrained relaxation of the H-positions using the MLIP, indicated by the suffix "(H constrained relaxation)". The main plot shows the distribution for the cases that result in a structural match, while the inset shows the results for those not resulting in a structural match. The full relaxation RMSD is normalized by the number of atoms, while in the other two cases it is normalized by the number of hydrogen sites.

### B. Structural agreement with DFT and EXP targets

Fig. S5 presents the distribution of the RMSD between the reference structure and the predicted one using the `pos-only-TD` model. The inset shows the distribution for the predictions that do not result in a match, while the main plot shows the distribution for the matching ones. The *Diffusion model* categories refer to structures that did not undergo any structural relaxation, that is, those that are directly taken from the output of the diffusion model. In case of the *H constrained relaxation*, the hydrogen sites were optimized using the `NequIP` MLIP. Similarly, the *Full relaxation* refers to a `NequIP` relaxation where all ionic positions are optimized.

First of all, the diffusion model output is already very close to the reference in many cases, both for the DFT and EXP datasets. In case of the EXP dataset, a wider tail is observed, which is expected to some extent. After performing the constrained relaxation of the hydrogen sites, the distributions can be significantly sharpened and pushed towards 0 Å. Important to note, the reference here are the structures from MC3D (either the `Quantum ESPRESSO` DFT relaxed ones or the EXP ones), i.e., they were not relaxed using the MLIP, in contrast to the comparisons in the main text. Thus, the deviations observed in this figure indicate the performance and deviations with respect to different methods, e.g., a different DFT code and settings, and still confirm a high agreement. This suggests the ability of the model to extend to different domains, in the sense that it can be used for crystal structures that are curated in different ways. Moreover, it is observed that the constrained relaxation tends to work slightly better for the `EXP` dataset, due to the different underlying methodologies outlined above, as it avoids that the host sites move too much. This trend is further confirmed for the `DFT` dataset.

## C. Benchmarking of MLIPs against DFT

Even though MLIPs typically achieve good performances nowadays, they should always be benchmarked for the specific applications. The following subsection discusses some benchmarking of various pretrained foundational MLIPs to select the most appropriate one for the present application. Each analyses compares 3 optimizers (`FIRE`, `BFGS`, `BFGS-linesearch`) that are available in the `ASE` [80] package, with 4 different values for the target relaxation threshold on 3 different datasets (`DFT`, `EXP`, `DFT-20-40`) that were discussed in the main text. Each of the data subsets contains around 100 selected structures from those cases that do not directly result in a structural match.

Fig. S6 starts with the discussion of the convergence rate (fraction of structures for which the final maximum force on an atom is smaller than *fmax*, when maximum 500 optimization steps are performed). Overall, most of the models manage to achieve convergence within 500 steps for almost all the structures. Only the `FIRE` algorithm struggles with very tight thresholds. This is also observed for the larger structures and the smallest threshold in case of the `BFGS-linesearch`. Among all the thresholds and datasets, the standard `BFGS` yields the most stable results.

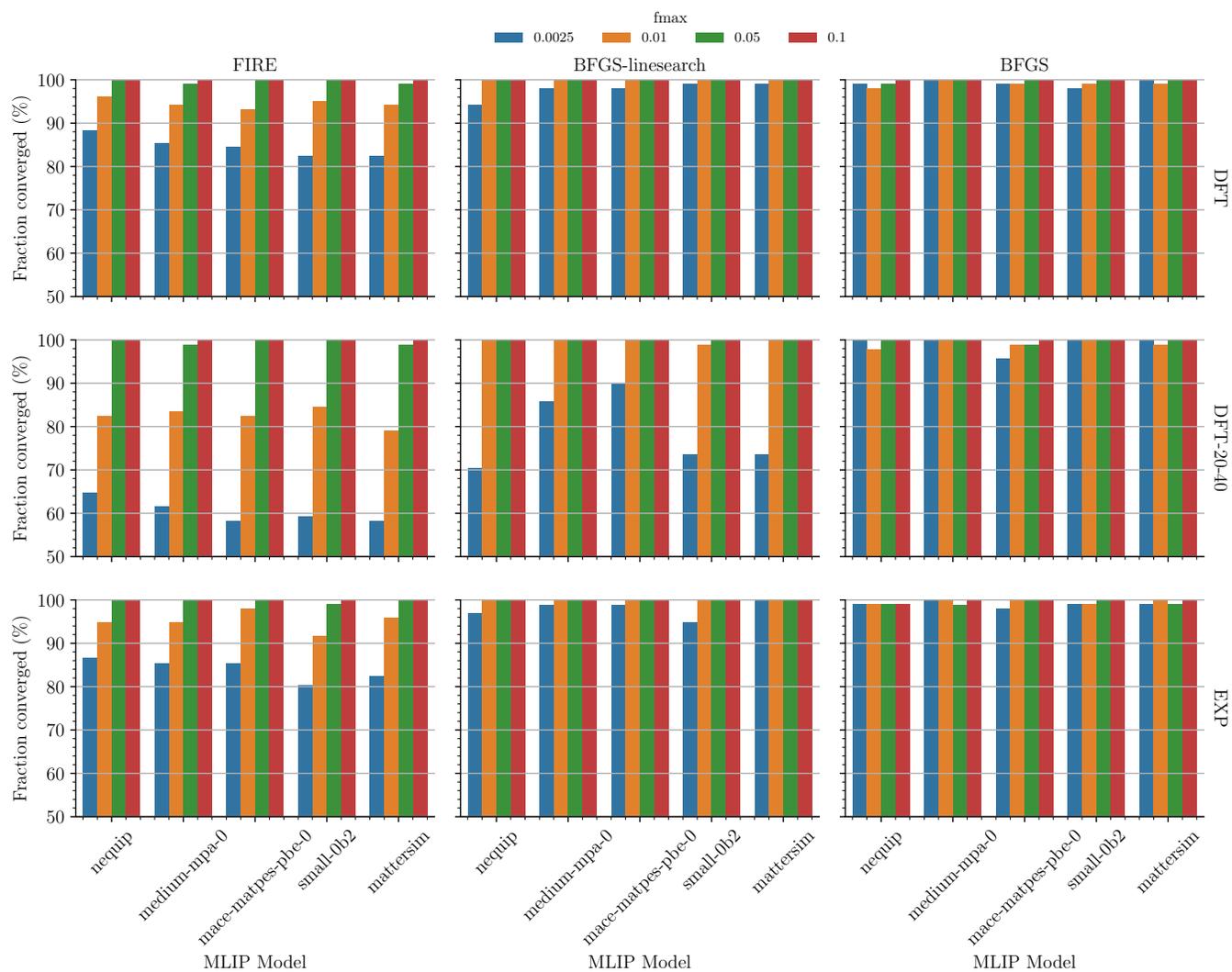

SUPPLEMENTARY FIG. S6. **MLIP relaxation convergence rate.** Percentage of successfully converged relaxations within 500 steps. The analysis compares 5 pre-trained foundational MLIPs across relaxations using three different optimizers, each of them with 4 different convergence thresholds *fmax* (in eV/Å). Each row presents the results for a different dataset: `DFT`, `DFT-20-40` and `EXP`.

Next, in Fig. S7 we discuss the MAE of the energy difference between the initial reference and the inpainted calculated with DFT and the MLIP. Important to note, we perform relaxations of the positions on the initial reference as well as on the inpainted one. It is especially important to understand how well the predicted energetic trends and differences agree between DFT and the MLIPs, with respect to the reference and prediction, as this was used in the main text (i) to select the most stable sample per structure and (ii) to determine the number of lower energy matches. First of all, one observes that the dataset can be ranked in ascending order with respect to the overall magnitude of the MAE as `DFT-20-40`, `DFT` and `EXP`. Again, and especially in case of the `EXP` dataset, the `BFGS` optimizer shows the best agreement. Moreover, significant differences across the different models can be observed in this benchmark. The `NequIP` model turns out to be the one that yields the best results. Finally, especially in case of the experimental structures, one also notices that the very tight threshold of 0.0025 eV/Å leads sometimes to worse results (most likely due to the larger noise as those structures are farther away from the (close to) equilibrium configuration that the models were mostly trained on). Based on this observation, the independence of the convergence rate with respect to $fmax$, and the fact that it also leads to better results in case of the experimental structures (with almost no effect on the other datasets), we will use $fmax = 0.01$ eV/Å in the final relaxations of this study. While the DFT relaxations were performed with a threshold of $fmax = 10^{-4}$ Ry/bohr $\approx 0.0025$ eV/Å, it is not required and—as shown—not beneficial to use the same strict thresholds for the MLIPs.

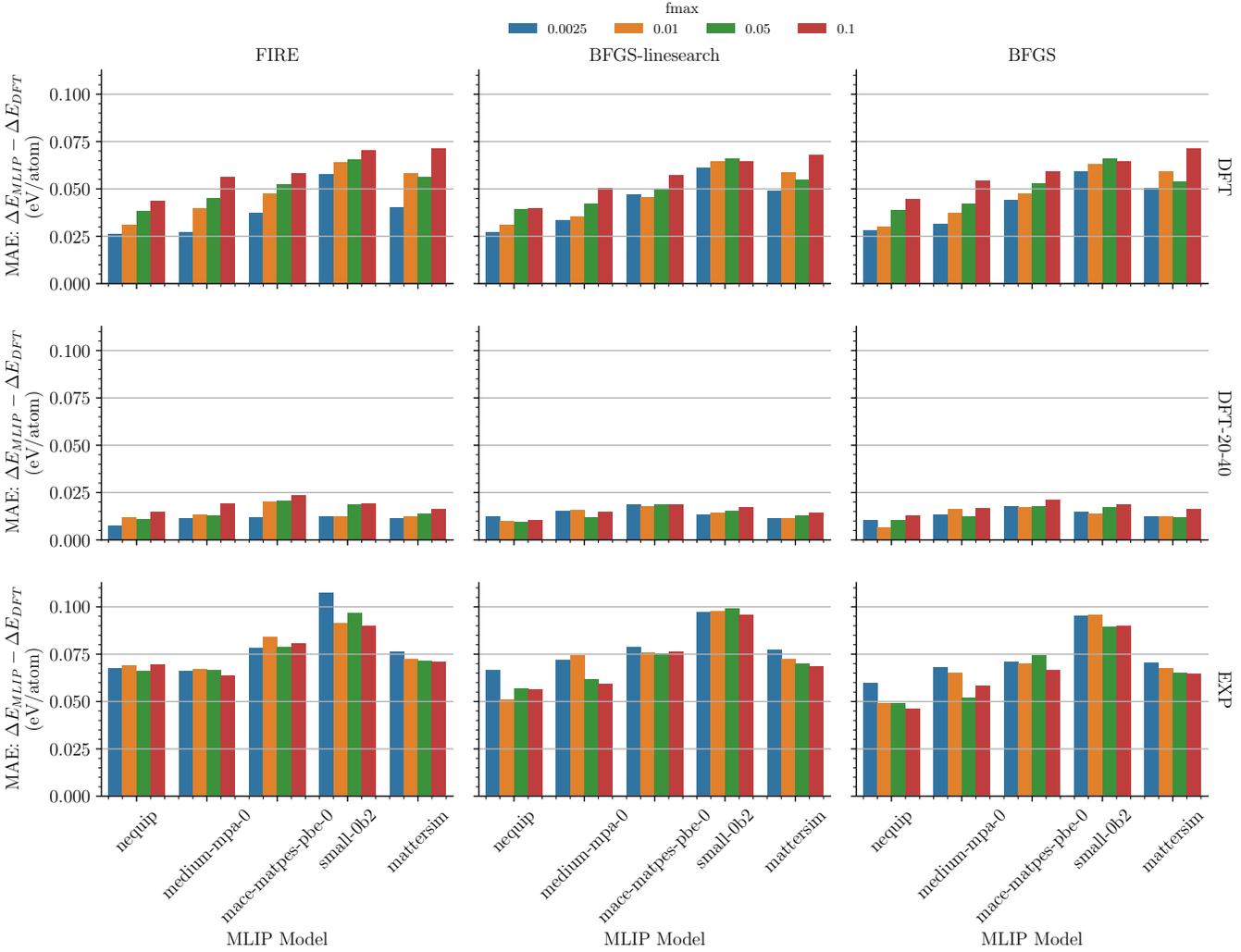

SUPPLEMENTARY FIG. S7. **MAE of the energy difference $\Delta E$ between the initial reference and predicted inpainted structure evaluated by MLIP and DFT.** The analysis compares 5 pre-trained foundational MLIPs across relaxations using three different optimizers, each of them with 4 different convergence thresholds $fmax$ (in eV/Å). Each row presents the results for a different dataset: `DFT`, `DFT-20-40` and `EXP`.

Finally, we discuss the previous aspect as a classification problem, which is even more in line with the application discussed in the main text. Fig. S8 shows the percentage of structures for which the stability prediction of the MLIP and DFT with respect to the energetic difference between the initial and inpainted structures agree, i.e., both predict the inpainted structure to be more stable or both classify it as less stable. This is exactly the information that was used in the main analysis.

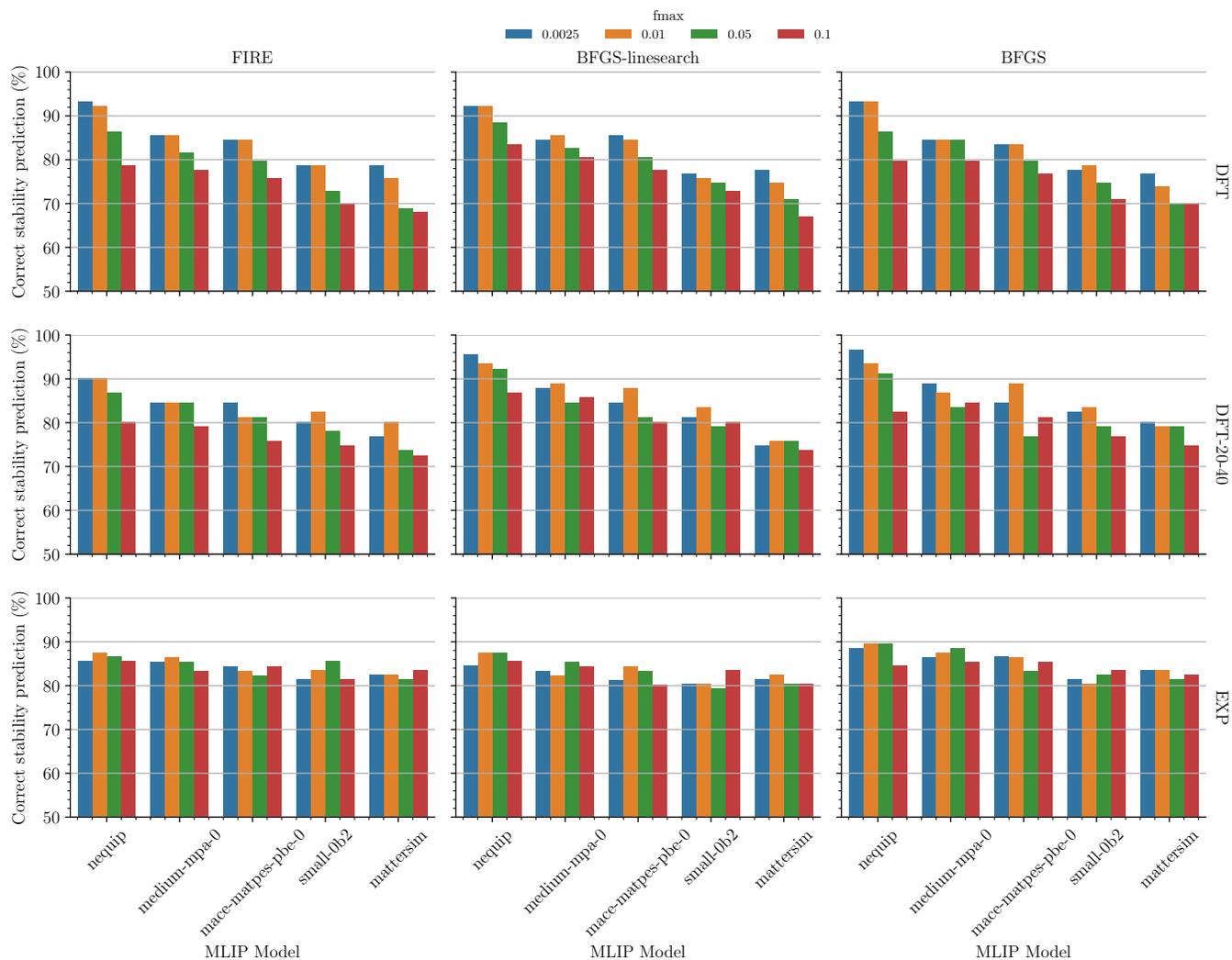

SUPPLEMENTARY FIG. S8. **Agreement of MLIP and DFT stability classifications.** Percentage of correct relative stability classifications, i.e., structures for which both DFT and the MLIPs predict the same sign of the energy difference $\Delta E$ between the initial reference and predicted inpainted structure. This measures in how many cases the MLIPs correctly classify whether the inpainted structure is more stable than the initial reference. The analysis compares 5 pre-trained foundational MLIPs across relaxations using three different optimizers, each of them with 4 different convergence thresholds $fmax$ (in eV/Å). Each row presents the results for a different dataset: `DFT`, `DFT-20-40` and `EXP`.

## D. Energetic ranking and performance impact

This section analyzes the energetic ranking of the samples for each structure that we obtain based on the MLIP, in our case `NequIP`, with respect to DFT. We select the 18 structures for this test for which neither a structural match nor a lower energy prediction is obtained for the `MLIP+DFT` dataset. As a reminder, we only discussed the sample with the lowest energy according to the MLIP in the main text. Here, we relaxed all the 30 samples for each of the 18 structures with DFT. Afterwards, we rank the samples by energy for each structure, with rank 0 corresponding to the minimum energy sample.

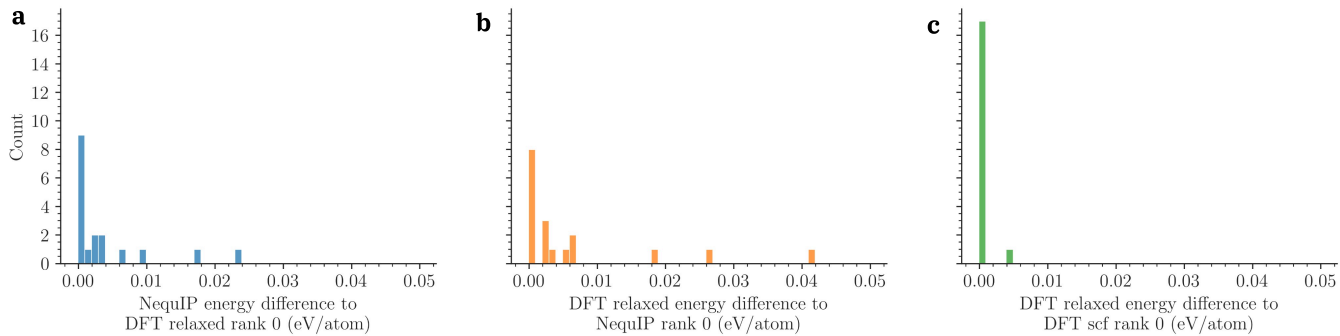

SUPPLEMENTARY FIG. S9. **Comparison of the energy ranking obtained with NequIP and DFT.** **a** `NequIP` energy relative to the rank 0 sample as selected by DFT relaxation. **b** DFT energy relative to the rank 0 sample as selected by `NequIP`. **c** SCF DFT energy relative to the rank 0 sample as selected by DFT after relaxation.

In Fig. S9a, we present the NequIP energy difference of the DFT rank 0 sample with respect to the NequIP rank 0 configuration. As can be seen in the figure, 9 out of the 18 structures show energy differences of less than 1 meV/atom. Hence, in those cases, the difference in the ranking is essentially due to numerical noise. On the other hand, some significant energetic differences according to NequIP are observed, indicating actual differences in the energetic ranking.

Similarly, we revert the previous analysis and analyze the energy difference based on DFT when selecting the NequIP rank 0 sample. In general, similar observations as before can be made, e.g., for many structures several samples relax to the same configuration and the different ranking is only due to numerical noise. However, more samples show energetic differences of a few meV/atom and 3 larger values are observed as well, indicating cases for which the sample selected based on NequIP is notably less stable according to our DFT results.

In summary, these analyses suggest that, in order to further improve the performance, the most reliable approach would be to run DFT calculations for all the samples. However, especially in a high-throughput context, running 30 relaxations per structure that one tries to reconstruct represents a significant computational cost. Fig. S9c compares the energy differences of the relaxed DFT structures with respect to the rank 0 sample when selecting it only based on a DFT SCF calculation, i.e., no structural relaxation. It can be seen that, except for 2 small differences of the order of 5 meV/atom, the SCF would already select the correct sample. In this way, one trade-off could be to do single-point energy evaluations based on DFT and only run the relaxation for the most stable one. Nonetheless, we highlight again the generally very good agreement of the NequIP energy ranking that allow to process a larger number of structures much faster.

In general, the spread across the samples for each structure, i.e., the difference between the highest and lowest energy is captured quite well by NequIP with respect to DFT, as shown in Fig. S10. One potential source of the wrong energetic ranking could be the known over-softening of MLIPs [81]. Nonetheless, one has to keep in mind that NequIP is a pretrained foundational model that was trained on a dataset with different DFT settings. Hence, finetuning could potentially resolve the differences as well.

Finally, taking the previous 18 structures, we also see that we can even further improve the performance of our approach. When selecting the rank 0 sample according to DFT, we can find 9 additional structural matches. Based on the previous samples selected by NequIP, no structural or lower energy match could be obtained. However, relaxing all samples by DFT reveals that the correct DFT minimum would result in a match (meaning that the `NequIP` relaxation ends up in another local minimum that is less stable). In two additional cases, a lower energy prediction could be obtained with respect to the original reference. Therefore, by evaluating everything with DFT, we could further increase our LES matching rate for the `DFT+MLIP` dataset to 99%. Similarly, just running a SCF DFT energy evaluation to select the most stable sample also results in 10 out of 18 additional LES matches.

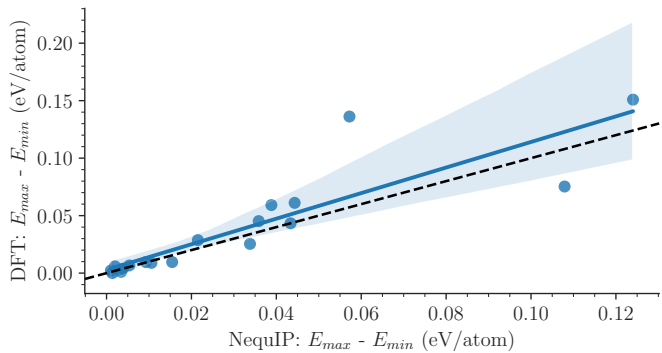

SUPPLEMENTARY FIG. S10. **Comparison of the energy spread obtained with NequIP and DFT.** DFT energy spread (difference between the maximum and minimum energy across the 30 samples per structure) vs. the energy spread evaluated by NequIP. All the samples have been relaxed with the corresponding methodology. The black dashed line represents the diagonal.

### E. Outlook: How to estimate the number of hydrogen sites to be added

In this section, we analyze a potential way of determining the number of hydrogen sites to be added. We randomly select 20 structures. For each structure, we generate samples with $N_{inpainted} \in \{1, \ldots, N_{max}\}$ added hydrogen sites. Moreover, for each $N_{inpainted}$, we generate 5 samples. Afterwards, we perform structural relaxations using `NequIP` and calculate the formation energies. For each of the 20 structures, we construct the phase diagram (formation energies as a function of composition) and calculate the energy above the resulting convex hull. Since the phase space is very sparsely covered, these results are of course not representing the actual stability. Nonetheless, Table S1 shows that the stability is indeed a good estimator to select the number of hydrogen sites to be added. In addition to the original composition, all stable compositions are listed (energy above the convex hull is 0 eV/atom, i.e., that sample is part of the convex hull). We consider this approach as successful when the correct composition is included in the stable compositions. As shown in the table, in most cases several samples are obtained. Nonetheless, this approach significantly reduces the number of compositions to be checked compared to the number of different initial compositions (given by $N_{max}$). In some cases, only the correct composition is labeled as stable. Overall, the approach correctly identifies the number of sites to be added (potentially among other candidates) in 17 out of 20 cases. This result is already promising and builds the foundation to be further extended in future work.

### F. Similarity with training set

This subsection briefly investigates the similarity of the training set with the test set, a very important aspect in order to accurately estimate the performance for actual applications. To quantify the similarity between the training set and the testing sets, in addition to removing overlapping compounds as described in section IV C in the main text, we follow the approach described by Mazitov *et al.* in Ref. [82]. We use the features of their `PET-MAD` MLIP and treat them as a structural fingerprint. The similarity is then evaluated by calculating the Euclidean distance of two fingerprints in that high-dimensional feature space.

Fig. S11 shows the similarity when comparing the `Alex-MP-20` training dataset with itself. This is done to have a reference of how similar structures are in a pre-filtered and unique training dataset. Moreover, it shows the smallest distance for each structure in `MC3D-DFT` and `MC3D-EXP` against all structures in `Alex-MP-20`. In Fig. S11b, the distribution of all pairwise distances is shown. Here, the different datasets show more similar distributions, with the main difference in the intensity of the peaks and broader tails in case of the `MC3D` datasets. This subplot is mainly intended to visualize the overall similarity and to sanity check the method with the findings reported in Ref. [82], where this similarity approach was presented and also applied to MC3D.
In Fig. S11a one observes that the distributions of the MC3D based test datasets are clearly shifted towards larger distances, compared to the already cleaned and filtered `Alex-MP-20` dataset. We use this as another indication that the performed tests are valid and that our data curation process was indeed careful.

SUPPLEMENTARY TABLE S1. **Proof of concept to determine the number of hydrogen sites in a given host structure.** We list the original composition to be determined, the stable compositions according to the formation energies when considering $N_{max}$ different compositions (with the number of hydrogen sites ranging from $1 \ldots N_{max}$) and whether the correct number of hydrogen sites $N_H$ is recovered, i.e., the composition is stable.

| Original Composition | Stable Compositions | $N_{max}$ | Recovered $N_H$ |
|---|---|---|---|
| $PH_2F_3$ | $P_2HF_6$, $PH_2F_3$, $PHF_3$ | 12 | True |
| $BeAlH_5$ | $BeAlH_5$ | 16 | True |
| $LiGaH_4$ | $LiGaH_4$ | 16 | True |
| $KH_4O_2F$ | $K_2H_3(O_2F)_2$, $K_2H(O_2F)_2$, $KHO_2F$, $KH_4O_2F$ | 12 | True |
| $CsAs(HO_2)_2$ | $CsAs(HO_2)_2$, $CsAsH_3O_4$, $CsAs(HO)_4$, $Cs_2As_2HO_8$ | 8 | True |
| $AlHSO_5$ | $Al_2H(SO_5)_2$, $AlHSO_5$ | 6 | True |
| $CaGaH_5$ | $CaGaH_5$, $CaGaH$, $Ca_2Ga_2H_3$ | 16 | True |
| $CuHOF$ | $CuHOF$, $Cu_2H(OF)_2$, $Cu_4H(OF)_4$, $Cu_4H_3(OF)_4$, $Cu_2H_3(OF)_2$ | 8 | True |
| $Na_2H_6PtO_6$ | $Na_2HPtO_6$, $Na_2H_6PtO_6$, $Na_2H_8PtO_6$, $Na_2H_5PtO_6$ | 11 | True |
| $MnNiHO_3$ | $Mn_2Ni_2HO_6$, $MnNiHO_3$ | 10 | True |
| $PHF_2$ | $P_4HF_8$, $PHF_2$, $P_2HF_4$ | 8 | True |
| $H_2SO_4$ | $HSO_4$, $H_2SO_4$, $H_3SO_4$, $H(SO_4)_2$, $H_3(SO_4)_2$ | 10 | True |
| $Dy(HO)_3$ | $DyH_2O_3$, $Dy(HO)_3$, $Dy_2HO_6$, $DyHO_3$ | 12 | True |
| $CsP(HO_2)_2$ | $Cs_2P_2HO_8$, $Cs_2P_2H_3O_8$, $CsP(HO_2)_2$ | 8 | True |
| $La_4H_9$ | $LaH_3$, $La_2H_5$, $La_4H_{11}$, $LaH_2$ | 16 | False |
| $ScH_3$ | $ScH_2$ | 17 | False |
| $K_2H_4Pt$ | $K_2H_2Pt$, $K_2H_6Pt$, $K_2H_4Pt$ | 14 | True |
| $HPbIO$ | $HPb_4(IO)_4$, $HPbIO$, $H_3Pb_4(IO)_4$ | 8 | True |
| $LiAlH_4$ | $LiAlH_4$ | 16 | True |
| $HC_2N_3$ | None | 10 | False |

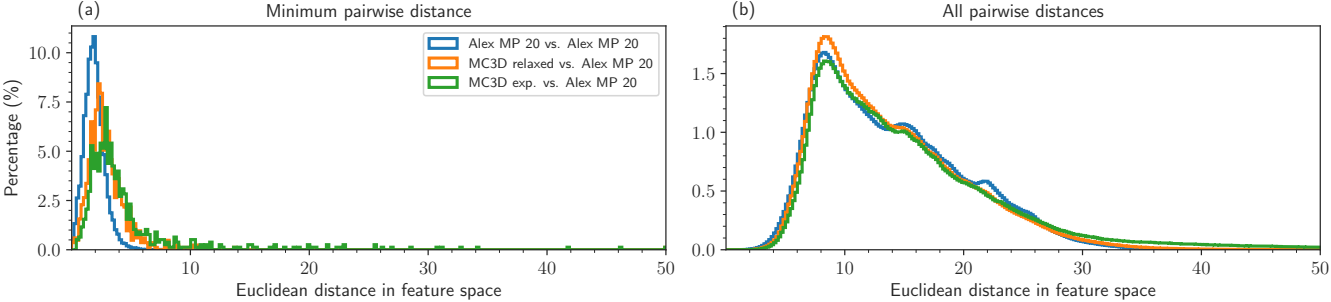

SUPPLEMENTARY FIG. S11. **Structural diversity and similarity of the training and testing datasets.** Euclidean distance of the 1024-dimensional features of the `PET-MAD` model [49] as a measure of structural diversity, as presented in [82]. **(a)** The minimum distance for each structure compared to all other structures in the reference dataset, **(b)** Distances for each structure compared to all other structures in the reference dataset. As a reference, the training dataset Alex-MP-20 is compared against itself, MC3D relaxed against Alex-MP-20 and the initial experimental structures from MC3D against Alex-MP-20. In all cases, we only consider structures that contain hydrogen.

## S2. SUPPLEMENTARY INFORMATION – DFT BASED RECONSTRUCTION APPROACH

### A. Underlying methodological idea

Hydrogen generally forms a covalent bond with one anion and may exhibit a weaker interaction with another, more distant, one. While the former implies a substantial charge redistribution, the latter, commonly referred to as a hydrogen bond, is mainly due to electrostatic and polarization effects that also include partial charge rearrangement [83, 84].

In both scenarios, a key information for developing an algorithm to restore the H atoms is the identification of electron density accumulating on the most electronegative atom(s) subsequent to the formation of the bond(s). A simple way to identify these elements is to introduce the electrons associated with the missing H atoms in a first principles simulation aimed at emphasizing the most probable candidate atoms for H to bind.

As a next step, rather than directly analyzing the charge density, the accuracy of the prediction can be improved by evaluating the electrostatic potential generated by the nuclei and the electrons in the aforementioned system, which immediately reveals electron-rich regions [85, 86]. The following quantity is thus used to formulate an algorithmic procedure aimed at identifying preferential hydrogen bonding sites:

$$V(\mathbf{r}) = \sum_A^N \frac{Z_A}{|\mathbf{R}_A - \mathbf{r}|} - \int \frac{\rho'(\mathbf{r}')\,\mathrm{d}\mathbf{r}'}{|\mathbf{r}' - \mathbf{r}|} \tag{S1}$$

where $N$ is the number of nuclei in the system, $\rho' = \rho - q/\Omega$ is the "balanced" charge density obtained by subtracting the constant background charge density $q/\Omega$ (with $\Omega$ unit cell volume) from the electronic density $\rho$ to compensate for the difference $q$ between the number of positive and negative charges in the system. The electrostatic potential has already extensively been used in the literature not only to identify H positions [86], but also to study the catalytic activity of nanoparticles [87] and more generally to study electrophilic and nucleophilic sites of molecules.

The method also takes inspiration from the approach that has been previously proposed to analyze Li motion in solid-state electrolytes but differs in many details that are described in the next section [51]. A similar approach has also been adopted for a high-throughput study aiming at identifying novel Li-ion cathode materials [88].

### B. Algorithmic procedure

The approach outlined in the previous section is turned into an iterative algorithmic procedure that starts with two inputs: the crystal structure of the material lacking hydrogen atoms in the crystallographic information provided, and the number of H atoms nominally present in the compound. The electrostatic potential in Eq. S1 is obtained from a plane-wave (PW) based DFT simulation of a negatively charged system with as many additional electrons as the missing hydrogen atoms. The PW basis set is an effective choice since $V(\mathbf{r})$ is only required in interstitial regions of the lattice where its description is not impaired by the use of pseudopotentials. Electrons accumulate on the most electronegative atoms and the tentative positions for placing hydrogen atoms are revealed by the maxima of $V(\mathbf{r})$.

The electrostatic potential is sampled on a fine grid and all the points that fall within a threshold of 99.5% of the highest value of $V(\mathbf{r})$ are identified as candidate H sites. A minimal allowed distance separating peaks in the electrostatic potential, set to 0.25 Å, is introduced in order to avoid placing H atoms too close to each other. An additional "proximity validation" is used to avoid placing H atoms on top of other atomic sites (in which case the algorithm fails). The electrostatic potential has, by definition, the same symmetry of the starting lattice and most often this procedure yields all symmetry equivalent positions for placing H, but it may happen that high symmetry points are not part of the grid. In that case the symmetry of the starting lattice is broken by the insertion of the H atom(s).

Depending on the number of maxima revealed by this first step, a decision is taken. When the available maxima can only accommodate a fraction of the hydrogen atoms to be restored, or when they match exactly, they are incorporated into the lattice. The position of the newly inserted atoms is then optimized with a relaxation of the H atoms, allowing the reorganization of the electronic density along the hydrogen bonds. It is worth mentioning that the structural relaxation is performed with a negatively charged cell accounting for the electrons of the hydrogen atoms that are yet to be restored and by keeping all the host atoms fixed, in order to preserve the initial symmetry and to avoid the dissociation of other bonds in the pristine lattice. The aforementioned procedure is reiterated with the new modified structure, now incorporating some of the hydrogen atoms to be reinstated.

Conversely, when the number of maxima exceeds the remaining hydrogen atoms to be restored in the structure, it is not possible to determine where to place them. This can happen for example when the multiplicity associated

with the candidate position is higher than the one of the correct position (due to a small displacement from a high symmetry point) or when the charge state of a group, for example $(NH_4)^+$ or $(OH)^-$, is not correctly identified. To alleviate these issues, the energy arising from the mutual interaction among the protons is added to the contribution resulting from their interaction with the electrostatic potential in Eq. S1. We will refer to this alternative final step as the *pinball* strategy. A combinatorial approach is used to compute the energy of all possible arrangements for hydrogen in the various positions identified from $V(\mathbf{r})$. While this problem grows like $\binom{n}{k}$ for a given number of $k$ H atoms and $n$ of maxima found for a given set of H, in the currently adopted scheme it requires a negligible amount of time for all cases considered in this work. After identifying the most energetically favorable configuration, similar to the previous case, the positions of H atoms are subsequently relaxed.

When all H atoms have been inserted, a final relaxation of the entire structure is performed.

The algorithm has been entirely built through the AiiDA infrastructure and written in Python. It is called `RestoreHydrogenWorkChain` and is included in the AiiDA plugin package "aiida-hydrogen-restorer", available on GitHub [74]. It leverages on the `PwBaseWorkChain`, already developed in the AiiDA framework, that interfaces with Quantum ESPRESSO [69, 70] submitting the self-consistent calculation, and on the `skimage.feature.peak` package, using the `peak_local_max` function to find the maxima of the electrostatic potential. For its validation we also use the `StructureMatcher` tool from the `pymatgen.analysis.structure_matcher` module to compare lattice structures.

### C. Computational details of the DFT based reconstruction approach

For all the computational investigations we used the scalar-relativistic pseudopotentials from PseudoDojo [89] version 0.4 with PBE as the exchange correlation functional and a plane-wave basis, implemented in Quantum ESPRESSO code. We set the energy convergence threshold `etot_conv_thr = 1E-05 Ry/atom` and force convergence threshold `forc_conv_thr = 1.0E-04 Ry/bohr` during the ionic minimization. The parameters generally follow the protocols used in MC3D [37], with the only exception of the smearing which is set to 0.01 Ry instead of 0.02 Ry.

### D. Results and discussion

The effectiveness of the algorithm was assessed by applying it to a set of structures obtained from the Materials Cloud three-dimensional crystals database (MC3D) [37] version `PBEsol-v1`, which is the PBEsol counterpart of the `DFT` dataset that was discussed in the main text. The algorithm presented here is not benchmarked on experimental structures.

For our study, we extracted all optimized hydrogen-containing structures, resulting in a set of 2909 structures. When running the `RestoreHydrogenWorkchain`, we encountered 116 failures unrelated to our algorithm (either due to an electronic convergence issue with Quantum ESPRESSO or node failures on the cluster). We subsequently removed those structures, obtaining a subset of 2793, henceforth referred to as the "test set". The total number of atoms per formula ranges up to 50. The percentage of hydrogen atoms per unit cell ranges up to 92%. The dataset contains different types of compounds, including organic minerals, composite crystals, and, by far the most abundant entry, inorganic crystals characterized by ionic, covalent or metal bonding. The structures were also categorized into subgroups based on their dimensionality, determined using the Low Dimensionality Structure Finder `lowdimfinder` [90–92] tool. This tool identifies sub-components within a periodic structure and provides their dimensionality. As expected, the predominant portion of the dataset comprises 3D structures, which are the focus of our study. Nonetheless, hydrogen is part of low dimensional fragments in 50% of the entries of the set. Other approaches [93–97] have already been developed for restoring hydrogen atoms in proteins and molecular systems and we expect these approaches to be easily extendable to low dimensional crystals or low dimensional fragments embedded in crystal voids. Nevertheless, we ran our algorithm on all structures, irrespective of their dimensionality, to verify the versatility of the proposed method.

In order to proceed with the validation of the algorithm, all hydrogen atoms are removed from the optimized structures and the iterative procedure described in the previous section is applied. Upon the successful completion of the hydrogen restoration algorithm, the total energy and the lattice structure obtained for each compound are compared with those obtained from the structural relaxation of the entry originally appearing in the "test set". We present the results in Fig. S12, distinguishing between cases that require the pinball procedure as the final step (dotted pattern) and those that do not (no additional pattern). On the $x$ axis, the compounds are divided into separate groups based on their dimensionality. In the histogram, green bars represent successful results. More specifically, light-green bars are used to represent the cases where the output structures match the input ones, while dark-green bars refer to structures that do not structurally match, but whose total energy is lower than the reference one. Red bars show instead structures that do not match and have higher total energy than the original ones. We start our discussion focusing on the structures that do not require the pinball method (hence, the success rates will be different from the

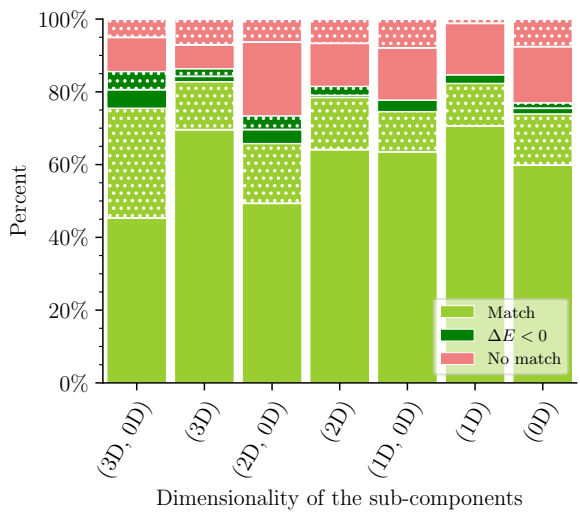

SUPPLEMENTARY FIG. S12. **LES matching rate based on the dimensionality of the structure (and of its sub-components).** Comparison of successful and failing structures, differentiated according to their dimensionality. The $x$ axis labels denote the different dimensions identified in the compounds. Structures that match are shown in light green, structures that do not match but have lower total energy than the reference one are in dark green, while those not matching and having a higher energy than the initial structure are in red. The dotted areas show the results that required the pinball strategy to complete.

ones seen in Fig. S12, as they are normalized with respect to all structures).

For zero-dimensional systems, i.e. molecular crystals, the success rate is 80%, which is understandably low, considering our focus on 3D systems. For 1D, 2D and 3D systems containing interstitial molecules (i.e. 0D components), a 82%, 72% and 84% success rate is respectively obtained. In this case, the main source of error is the charge state of interstitial molecules, and the presence of multiple relative minima in the potential energy surfaces, as will be discussed below. The accuracy increases when zero-dimensional systems are absent in the compounds. For 1D, 2D and 3D compounds we obtain 84%, 85% and 92% accuracy, respectively. In summary, a success rate of 87% is obtained on the subset of structures that do not require the application of the pinball method.

Overall, 76% of the structures did not require the application of the pinball method. We continue the discussion with the remaining 24% that required the pinball method. In the presence of 0D components, a success rate of 58%, 76% and 88% is obtained for 1D, 2D and 3D, respectively. When 0D components are absent, a success rate of 91%, 72% and 68% is found for 1D, 2D and 3D, respectively. The pinball scheme achieves an overall success rate of 71% on the subset of structures that required its application. Surprisingly, the pinball method seems to perform well even in the presence of low-dimensional structures, but the statistical analysis is hindered by the small size of these sets.

Combining the two aspects, as in Figure S12, we observe a success rate of 78%, 73% and 86% for 1D, 2D and 3D systems in the presence of 0D components, respectively. As discussed before, these success rates increase in the absence of 0D components to 85%, 82% and 86% for 1D, 2D and 3D systems, respectively. In summary, the algorithm achieves a success rate of 83% on the whole test set.

Further insight into the algorithm outcomes is provided by the total energy difference per atom for the 3D structures subset, $\Delta E$, defined as: $\Delta E = (E_{fin} - E_{in})/n$, where $E_{fin}$ represents the energy of the structure as output of the algorithm, $E_{in}$ is the energy of the structure used as input and previously geometrically optimized, and $n$ is the total number of atoms per crystal cell. The distribution of $\Delta E$ is shown in Fig. S13. We note that both matching and non-matching structures, represented by green and red bars respectively, extend well below and above $\Delta E = 0$. A broad, asymmetric distribution is observed for both cases, with the matching results piling up at $\Delta E = 0$ but also extending up to $\Delta E \sim -230$ meV/atom. Non-matching structures are spread out in the positive part of the $x$ axis up to $\Delta E \sim 850$ meV/atom, but also, as already anticipated, for many of them the total energy of the newly identified structure is significantly lower than the one of the reference system, extending up to -950 meV/atom.

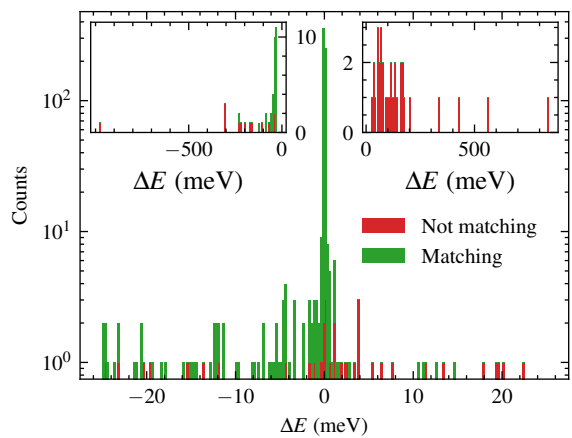

SUPPLEMENTARY FIG. S13. **Energetic comparison of the predictions and references.** Difference between the total energy $E_{in}$ for the original structure appearing in the test set and the final one $E_{fin}$ obtained from the hydrogen restoration algorithm, only considering the 3D compounds, for which all the atoms positions were relaxed. The green (red) bars display matching (non-matching) input and hydrogen-restored output structures.

### E. The challenge related to a greedy approach

Here, we present one example, $CrH_2$ [99], where the algorithm performs two steps placing one H at a time, first using electrostatic maxima (Fig. S14), then using the pinball method for the second H atom. This example highlights the challenges related to the greedy approach of the DFT-based algorithm. The final result is shown in the inset of Fig. S14, where our algorithm identifies the central position along the diagonal of the rhombohedral cell as optimal for placing the first H atom, thus spoiling the next step performed with the pinball strategy. The optimal atomic coordinates for placing two H atoms correspond instead to the two relative maxima of the electrostatic potential shown in the same figure obtained at the first iteration. The lack of inter-hydrogen interaction is clearly responsible for the mismatch and the resulting structure with a total energy 166.53 meV/atom higher than the experimental one. In fact, applying the pinball method from the beginning and suggesting as positions for the H insertion the three maxima in the potential shown in fig. S14, this results in the correct placement of the two H at [0.25, 0.25, 0.25] and [0.75, 0.75, 0.75]. While this example suggests that inter-hydrogen interactions should be included from the very beginning, the combinatorial explosion described above makes this strategy technically challenging. Moreover, it highlights the challenge related to the greedy nature of the algorithm and visualizes the advantage of the "global" approach as highlighted in the main text.

In summary, also the DFT based approach, which is more physically motivated, and has been used as a benchmark to validate the deep-learning based approach discussed in the main text, achieves quite good results in terms of structural matching rate, including identification of structures more stable than the original reference. The DFT based algorithm also performs well on the target lattice structures, i.e. 3D inorganic compounds (which do not require the pinball method), where it reaches 87% accuracy, but also shows good performance on low dimensional structures,

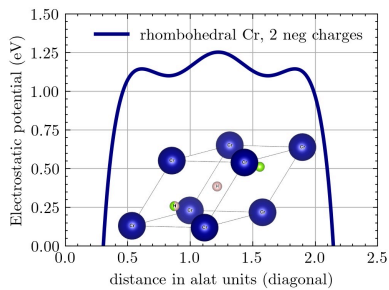

SUPPLEMENTARY FIG. S14. **Illustrating the challenges of the greedy DFT-based approach.** The electrostatic potential along the diagonal of the rhombohedral crystal cell of $CrH_2$, shown in the inset, used in the first step of the algorithm. The crystal structure image presented here was generated using VESTA [98].

with an overall accuracy of 83% on the materials composing our test set, thus suggesting the good versatility of the proposed algorithm. Further work would be required to fine tune the method especially concerning the orientation of interstitial molecules and a more gradual addition of inter-hydrogen interaction, but is beyond the scope of this work, also considering the extremely high success rate of the deep-learning based approach presented in the main text.



\end{document}